\documentclass[11pt,a4paper]{article}
\usepackage{amsmath,amssymb}
\usepackage{lipsum}
\usepackage{enumitem}
\usepackage{soul}
\usepackage{graphicx}
\usepackage{color}
\usepackage{slashed}
\usepackage{bbold}
\usepackage{wasysym}

\usepackage{multirow}

\usepackage[sort&compress,square,numbers]{natbib}

\usepackage{xspace}

\newcommand{\TeV}{{\ensuremath\rm TeV}\xspace}
\newcommand{\GeV}{{\ensuremath\rm GeV}\xspace}

\newcommand{\lb}{\left(}
\newcommand{\rb}{\right)}
\newcommand{\fb}{\ensuremath\rm fb}
\newcommand{\pb}{\ensuremath\rm pb}
\newcommand{\ab}{\ensuremath\rm ab}

\newcommand{\HSv}[1]{\texttt{HiggsSignals-#1}}
\newcommand{\HBv}[1]{\texttt{HiggsBounds-#1}}

\newcommand{\eqn}{equation}
\newcommand{\lam}{\lambda}

\def\mev{\;\hbox{MeV}}
\def\gev{\;\hbox{GeV}}

\graphicspath{ {./plots/} }

\begin{document}
\rightline{RBI-ThPhys-2020-55}
\vspace{8mm}

\begin{center}
{\LARGE \textbf{ IDM benchmarks for the LHC and future colliders}}
\\ [1cm]
{\large {Jan Kalinowski,$^{a}$ Tania Robens,$^{b}$ Dorota Soko\l owska$^{a,c}$\\ and Aleksander Filip \.Zarnecki$^{a}$
}}
\\[1cm]
\end{center}

\hspace{1cm}
\begin{minipage}{14cm}
\textit{ 
\noindent $^a$ Faculty of Physics, University of Warsaw, 
ul.~Pasteura 5, 02--093 Warsaw, Poland\\[0.1cm]
$^b$ Theoretical Physics Division, Rudjer Boskovic Institute,
10002 Zagreb, Croatia\\[0.1cm]
$^c$ International Institute of Physics, Universidade Federal do Rio Grande do Norte,
Campus Universitario, Lagoa Nova, Natal-RN 59078-970, Brazil
}
\end{minipage}

\vspace{0.7cm}

\begin{abstract}
We present cross-section expectations for various processes and collider options, for {benchmark scenarios of} the Inert Doublet Model, a Two Higgs Doublet Model with a dark matter candidate. The proposed scenarios are consistent with current {dark matter} constraints, including the most recent bounds from the XENON1T experiment and relic density, as well as with known collider and low-energy limits. These benchmarks, chosen in earlier work for studies at $e^+e^-$ colliders, exhibit a variety of kinematic features that should be explored at current and future runs of the LHC. We provide cross sections for all relevant production processes at 13 \TeV, 27 \TeV and 100 \TeV proton collider, as well as for a possible 10 \TeV and 30 \TeV muon collider. 
\end{abstract}

\def\thefootnote{\arabic{footnote}}
\setcounter{footnote}{0}

\section{Introduction}

The LHC discovery of a scalar particle consistent with the Standard Model (SM) predictions left many questions unanswered, among which is the lack of a dark matter candidate. This motivates investigations of beyond the SM extensions  of the scalar sector.  
The Inert Doublet Model (IDM) \cite{Deshpande:1977rw,Cao:2007rm,Barbieri:2006dq},  a Two Higgs Doublet Model with a discrete $Z_2$ symmetry, is a simple and well motivated model that leads to a {stable} dark matter candidate. It has been discussed widely in the literature (see e.g. \cite{LopezHonorez:2006gr,Honorez:2010re,Dolle:2009fn,Goudelis:2013uca,Lundstrom:2008ai,Dolle:2009ft,Gustafsson:2012aj,Aoki:2013lhm,Ho:2013spa,Arhrib:2013ela,Arhrib:2012ia,Swiezewska:2012eh,Krawczyk:2013jta,Ginzburg:2014ora,Belanger:2015kga,Blinov:2015qva,Ilnicka:2015jba,Hashemi:2015swh,Arhrib:2015hoa,Poulose:2016lvz,Datta:2016nfz,Hashemi:2016wup,Kanemura:2016sos,Akeroyd:2016ymd,Biondini:2017ufr,Wan:2018eaz,Ilnicka:2018def,Belyaev:2018ext,Kalinowski:2018ylg,Kalinowski:2018kdn,Dercks:2018wch,Bhardwaj:2019mts,Kanemura:2019kjg,Banerjee:2019luv,Lu:2019lok,Braathen:2019pxr,Braathen:2019zoh,Guo-He:2020nok,Abouabid:2020eik,Fabian:2020hny,Basu:2020qoe,Jangid:2020qgo,Banerjee:2021oxc,Banerjee:2021anv,Banerjee:2021xdp,Banerjee:2021hal,Yang:2021hcu}), and we refer the reader to this discussion for further reference.

The {imposed} discrete $Z_2$ symmetry (called $D$-symmetry) {corresponds to} the following transformation properties:
\begin{equation}\label{eq:symm}
\phi_S\to \phi_S, \,\, \phi_D \to - \phi_D, \,\,
\text{SM} \to \text{SM}.
\end{equation}
Here the $\phi_S$ doublet plays the same role as the analogous doublet in the Standard Model, {providing} the SM-like Higgs particle. This doublet is even under the $D$-symmetry, while the second doublet, the inert (or dark) $\phi_D$, is $D$-odd and contains four scalars, two charged and two neutral ones, labelled $H^\pm$ and $H,A$, respectively. In the rest of this work, we consider cases where $H$ serves as the dark matter candidate of the model.

 We refer here to our previous analysis \cite{Kalinowski:2018ylg,Kalinowski:2018kdn,deBlas:2018mhx}, where we proposed benchmark scenarios with an emphasis on the discovery potential at $e^+e^-$ colliders. The benchmarks presented in this work were chosen to cover a large range of the parameter space {relevant at colliders}, especially regarding the mass differences in the dark scalar sector. {In particular, we divided the benchmark points into two categories, roughly split into areas where the new scalar masses are below 300 \GeV or reach up to 500 \GeV. As mass spectra are usually relatively degenerate for these particles \cite{Ilnicka:2015jba,Ilnicka:2018def,Kalinowski:2018ylg}, especially for higher masses $\gtrsim\,300\,\GeV$, all scalar masses are relatively close, so a characterization by one scale is sufficient. For lower mass scales, the dark matter candidate can be lighter than the unstable scalars masses. Another important point is the on- or off-shellness of the decay products, which in this case are electroweak gauge bosons. As major backgrounds stem from the production of such bosons, together with missing energy, such features are an important selection criterium for signal over background enhancement. In total, we consider 40 specific parameter points, split into the low and high mass regions as discussed above. A more detailed description of the specific characteristics of these benchmark points is given in section \ref{sec:benchmarks} below.} 

 In this work cross-section predictions are given for these benchmarks, for a variety of production processes at the 13 and 27 \TeV~LHC, for a 100 \TeV proton-proton collider, as well as for a muon collider. The unstable dark scalars decay as $A\,\rightarrow\,H Z\,(100\%)$ and $H^\pm\rightarrow\,W^\pm H$ (dominantly) for all points considered, where the above decays can be on- or off-shell depending on the mass spectra. Cross sections were calculated using Madgraph5 \cite{Alwall:2011uj} with a UFO input file from \cite{Goudelis:2013uca}\footnote{\label{foot:ufo} Note the official version available at \cite{ufo_idm} exhibits a wrong CKM structure, leading to false results for processes involving electroweak gauge bosons radiated off quark lines. In our implementation, we corrected for this. Our implementation corresponds to the expressions available from \cite{Zyla:2020zbs}.}.

In order to assess the possible collider reach, we then resort to a very simple counting criterium, and mark a benchmark point as reachable if at least 1000 events will have been produced for a specific collider scenario, using the colliders nominal center-of-mass energy and design luminosity. We acknowlegde that this simple comparison criterium can only serve as a first step, and needs to be further tested by including full signal and background simulation, including the development of specific search strategies. However, we find this useful to provide first guidance for the benchmark points considered here.

The IDM is distinct in the sense that its unique signatures are mostly SM electroweak gauge boson and missing (transverse) energy\footnote{VBF-type SM scalar production with invisible decays in the off-shell mode is also an important channel, cf. e.g. \cite{Dercks:2018wch}.}. As couplings in both electroweak production and decay are determined by SM parameters (see e.g. discussion in \cite{Ilnicka:2018def}), rate predictions depend on a very small number of new physics parameters, typically mainly the masses of the new scalars; we will give examples to exceptions to this in the main body of this manuscript. This distinguishes it from other scalar extensions where a large number of additional parameters plays a role. While production modes can be similar to standard two Higgs doublet models, the exact $Z_2$ symmetry prevents couplings of the new scalars to fermions and therefore leads to distinct signatures of electroweak gauge bosons and missing (transverse) energy. \\
Finally, we want to briefly comment on other new physics models that lead to similar final states. In particular, many searches have been carried out by the LHC experiments within supersymmetric frameworks, cf. e.g. \cite{atlas_susy,cms_susy}. Supersymmetric models can also lead to multilepton signatures and missing transverse energy. In \cite{Belanger:2015kga,Dercks:2018wch}, recasts of such searches within the IDM were considered. The parameter space in \cite{Belanger:2015kga} which is excluded by LHC Run 1 searches is however equally excluded by dark matter considerations, as it features quite low dark matter masses which would lead to an overclosure of the universe. In \cite{Dercks:2018wch}, a heuristic argument was given why multilepton SUSY searches tend to cut out parameter regions in the IDM that would a priori lead to high event rates. 
Another model one could consider in this respect is the THDMa \cite{Ipek:2014gua,No:2015xqa,Goncalves:2016iyg,Bauer:2017ota,Tunney:2017yfp,Abe:2018bpo,Robens:2021xgz}, a two Higgs Doublet model with an additional pseudoscalar that, in the gauge eigenstate, serves as a portal to a dark sector. Again, dilepton and missing transverse energy signatures are one of the prime channels of this model. However, both this and the SUSY scenarios come with topologies different from the one which lead to these final states in the IDM. A more detailed comparison of the consequences of these differences is in the line of future work.

\section{The IDM \label{sec:model}}

The scalar sector of the IDM consists of two {SU(2)$_{L}$} doublets of complex scalar fields,   $\phi_S$  and $\phi_D$, {{with the $ D$-symmetric potential:}}
\begin{equation}\begin{array}{c}
V=-\frac{1}{2}\left[m_{11}^2(\phi_S^\dagger\phi_S)\!+\! m_{22}^2(\phi_D^\dagger\phi_D)\right]+
\frac{\lambda_1}{2}(\phi_S^\dagger\phi_S)^2\! 
+\!\frac{\lambda_2}{2}(\phi_D^\dagger\phi_D)^2\\[2mm]+\!\lambda_3(\phi_S^\dagger\phi_S)(\phi_D^\dagger\phi_D)\!
\!+\!\lambda_4(\phi_S^\dagger\phi_D)(\phi_D^\dagger\phi_S) +\frac{\lambda_5}{2}\left[(\phi_S^\dagger\phi_D)^2\!
+\!(\phi_D^\dagger\phi_S)^2\right].
\end{array}\label{pot}\end{equation}
Exact $D$-symmetry (cf. eq. (\ref{eq:symm})) implies that only  $\phi_S$ can acquire  {a} nonzero vacuum expectation value ($v$). As a result the  scalar field{s} from different doublets  do not mix, and {the} lightest particle from $\phi_D$ is stable.  
{The dark sector  {contains} four new particles: $H$, $A$ and $H^{\pm}$.  We here choose $H$ to denote the dark matter candidate (choosing $A$ instead is equivalent to changing the sign of $\lambda_{5}$). 

The model contains seven free parameters after electroweak symmetry breaking. The SM-like Higgs mass $M_h$ and the vev, $v$, are fixed by LHC measurements as well as electroweak precision observables. We choose the remaining five free parameters to be
\begin{\eqn}\label{eq:physbas}
M_H, M_A, M_{H^{\pm}}, \lam_2, \lam_{345},
\end{\eqn}
 where  {the} $\lambda$'s refer to coupling{s} within the dark sector and to the SM-like Higgs, respectively, with $\lam_{345} = \lam_{3}+\lam_{4}+\lam_{5}$.

\section{Experimental and theoretical constraints \label{sec:constraints}}

We consider the following experimental and theoretical constraints on the model (see e.g. \cite{Ilnicka:2015jba,Kalinowski:2018ylg} for a more detailed discussion):

\begin{itemize}

\item{Positivity constraints}: we require that the potential is bounded from
 below.

\item{Perturbative unitarity}: we require the scalar {$2\,\rightarrow\,2$}
 scattering matrix to be unitary.

\item{Global minimum}: in the IDM two neutral minima can coexist 
even at tree level. Unless the following relation is satisfied
\begin{equation}\label{eq:invac}
\frac{m_{11}^2}{\sqrt{\lam_1}}\,\geq\,\frac{m_{22}^2}{\sqrt{\lam_2}},
\end{equation}
the inert minimum is only a local one, with the global vacuum 
corresponding to the case of massless fermions \cite{Ginzburg:2010wa}.
We impose the above relation in our scan.

\item{Higgs mass and signal strengths}: the mass of the SM-like 
Higgs boson $h$ is set to $$M_h=125.1\,\gev \label{Eq:mhexp},$$ in 
agreement with limits from ATLAS and CMS experiments \cite{Aaboud:2018wps,Mondal:2018hxy}, 
while the total width of the SM-like Higgs boson obeys an upper
 limit of  \cite{Sirunyan:2019twz}
\begin{\eqn}\label{eq:gtot}  
\Gamma_\text{tot}\,\leq\,9\,\mev.
\end{\eqn}

We have confirmed that all points obey the newest limit for invisible Higgs boson decays, $\text{BR}_{h\,\rightarrow\,\text{inv}}\,\le\,0.15$ \cite{ATLAS-CONF-2020-052}.
Furthermore, all points have been checked against currently available signal strength measurements, including simplified template cross-section information, using the publicly available tool \HSv2.6.0 \cite{Bechtle:2013xfa,Bechtle:2020uwn}, where we require agreement at $95\%$ confidence level.

\item{Gauge bosons width}: introduction of light new particles could in principle significantly change the total width of electroweak gauge bosons {(cf. e.g. \cite{Zyla:2020zbs})}. To ensure that $W^\pm \to H H^\pm$ and $Z\to HA,H^+H^-$ decay channels are kinematically forbidden we set:
\begin{equation}\label{eq:gwgz}
M_{A,H}+M_{H^\pm}\,\geq\,M_W,\,M_A+M_H\,\geq\,M_Z,\,2\,M_{H^\pm}\,\geq\,M_Z.
\end{equation} 

\item{Electroweak precision tests}: we call for a $2\,\sigma$ (i.e. $95 \%$ C.L.) agreement with electroweak precision observables, parametrized through the electroweak oblique parameters $S,T,U$ \cite{Altarelli:1990zd,Peskin:1990zt,Peskin:1991sw,Maksymyk:1993zm}, {tested against the latest results from the GFitter collaboration \cite{gfitter,Haller:2018nnx}}. In our work, calculations were done through the routine implemented in the Two Higgs Doublet Model Calculator (\texttt{2HDMC}) tool \cite{Eriksson:2009ws}, which checks whenever model predictions fall within the observed parameter range.

\item{Charged scalar mass and lifetime}: we take a conservative lower estimate on the mass of $M_{H^\pm}$ following analysis in \cite{Pierce:2007ut} to be 
\begin{\eqn}\label{eq:mhplow}
M_{H^\pm}\,\geq\,70\,\gev.
\end{\eqn}
We also set an upper limit on the charged scalar lifetime of 

\begin{\eqn}\label{eq:limcharged}
\tau\,\leq\,10^{-7}\,s,
\end{\eqn}
in order to evade bounds from quasi-stable charged particle searches\footnote{More detailed studies using recasts of current LHC long-lived particle searches can be found in \cite{Heisig:2018kfq,Belyaev:2020wok}.}.

\item{Collider searches for new physics}: we require agreement with the  null-searches from the LEP, Tevatron, and LHC experiments. We use the publicly available tool \HBv5.9.0 \cite{Bechtle:2008jh, Bechtle:2011sb, Bechtle:2013wla,Bechtle:2015pma,Bechtle:2020pkv}. In addition
the reinterpreted LEP II searches for supersymmetric particles analysis exclude the region of masses in the IDM where simultaneously \cite{Lundstrom:2008ai}
\begin{equation}\label{eq:leprec}
M_A\,\leq\,100\,\gev,\,M_H\,\leq\,80\,\gev,\,\, \Delta M {(A,H)}\,\geq\,8\,\gev,
\end{equation}
as it would lead to a visible di-jet or di-lepton signal. After taking into account all the above limits we are outside of the region excluded due to the reinterpretation of the supersymmetry analysis from LHC Run I \cite{Belanger:2015kga}.

\item{Dark matter phenomenology}: we apply dark matter relic density limits obtained by the Planck experiment \cite{Aghanim:2018eyx}:
\begin{equation}\label{eq:planck}
\Omega_c\,h^2\,=\,0.1200\,\pm\,0.0012.
\end{equation}
For a DM candidate that provides 100\% of observed DM in the Universe we require the above bound to be fulfilled within the 2$\sigma$ limit. However, we also allow for the case where $H$ is only a subdominant DM candidate, with 
\begin{\eqn}\label{eq:om_limapp}
\Omega_H h^2 < \Omega_c\,h^2
\end{\eqn}
Note that this also leads to a rescaling of the respective direct detection limits \cite{Ilnicka:2015jba,Kalinowski:2018ylg}.

In the results presented here, we apply XENON1T limits \cite{Aprile:2018dbl}\footnote{We use a digitized format of that data available from \cite{PhenoData}.}. {For consistency, we here calculated the dark-matter related variables using \texttt{micrOMEGAs$\_$5.0.4} \cite{Belanger:2018mqt} \footnote{{Note that for some points, relic density values change using the most up-to-date version, i.e. \texttt{micrOMEGAs$\_$5.2.4} \cite{Belanger:2020gnr}. Similar results can be obtained by changing the integration mode for some points. We list the corresponding values for the low-mass benchmark points in appendix \ref{app:newmo} for reference. }}}.

\end{itemize}

\subsection{Requiring exact relic density}

As discussed above, we here require relic density to be below the current value as determined by the Planck collaboration (cf. eq. (\ref{eq:planck})). In the Inert Doublet model, meeting the exact relic density is only possible in certain mass ranges. We here enhance a previous discussion on this which was presented in \cite{Kalinowski:2018ylg} (see also the discussion in \cite{Belyaev:2016lok}).
\begin{itemize}
\item{}{\bf Lower bound on dark matter mass}\\
A combination of signal strength measurements for the 125\,GeV resonance sets an upper limit on the absolute value of the coupling $\lam_{345}$, which determines the $H\,H\,h$ coupling. In this area, the major annihilation channel is $H\,H\,\rightarrow\,b\,\bar{b}$, mediated by $h$-exchange. Low values of $\lam_{345}$ in turn lead to large values for the relic density, as annihilation cross sections are taking lower values. In principle, co-annihilation with $A$ or $H^\pm$ could remedy this, leading to larger annihilation cross sections, for mass splittings which are smallish. Indeed, in \cite{Belyaev:2016lok} this scenario is explicitly discussed (see also \cite{Dolle:2009fn}). The combination of these bounds leads to a lowest value of $M_H\,\sim\,55\,\GeV$ \cite{Ilnicka:2015jba,Belyaev:2016lok,Ilnicka:2018def,Dercks:2018wch}. In a more detailed scan, however, we find that masses can in principle be as low as around $44\,\GeV$, if the mass difference between $M_A$ and $M_H$ is quite small, up to $4\,\GeV$; {the} dominant contribution then comes from coannihilation of $HA\,\rightarrow\,{\lb d\,\bar{d},\,s\,\bar{s},\,b\,\bar{b}\rb}$, but none of these points results in the correct relic density.
\item{}{\bf Resonance region, $M_H\,\sim\,M_h/2$}\\
In this region, the main annihilation channels are $h$-mediated, primarily into $b\bar{b}$ {and $W^+W^-$} final states. This leads to points that meet the exact relic density, with smallish {$|\lam_{345}|\,\lesssim\,0.006$} values. 
\item{}{\bf Region up to around $75\,\GeV$}\\
In this region, $HH$ anihilation into (partially off-shell) $W^+ W^-$ final states start to dominate. Due to interference effects between $h$-mediated and quartic couplings (see e.g. \cite{Honorez:2010re,Belyaev:2016lok}), some points in the mass range around $70-73\,\GeV$ render exact relic density, including all current constraints. Absolute values for $\lam_{345}$ are $\lesssim\,0.006$ in that region. {As in the low mass region, for quite mass-degenerate scalars, $M_H-M_A\,\lesssim\,7\,\GeV$, the dominant annihilation process is given by $HA\,\rightarrow\,\lb d\,\bar{d},\,s\,\bar{s},\,b\,\bar{b} \rb$; none of these points however renders the correct relic density.}
\item{}{\bf Region between $75\,\GeV$ and $160\,\GeV$}\\
This region was proposed in \cite{LopezHonorez:2010tb} as a good region for dark matter relic density in the IDM, where the calculation depends on cancellations between diagrams for $VV^{*}$ final states. However, the values for $\lam_{345}$ required here are by now ruled out by limits from direct detection experiments.
{The dominant annihilation channel is $H\,H\,\rightarrow\,W^+\,W^-$.}

\item{}{\bf Region between $160\,\GeV$ and around $500\,\GeV$}\\
In this region, currently no study exists that provides scenarios within the IDM where exact relic density can be generated. Examples for studies are given in \cite{Goudelis:2013uca,Arhrib:2013ela,Ilnicka:2015jba}. Largest values of relic density stem from $HH\,\rightarrow\,W^+ W^-$ annihilation, with annihiliation rates too large to render the exact value.
\item{}{\bf Larger masses, $M_H\,\gtrsim\,500\,\GeV$}.\\
Here, the exact values of relic density can be obtained if mass splittings between dark scalars are quite small, roughly $\lesssim\,{10}\,\GeV$ (see also discussion in \cite{Hambye:2009pw}). The dominant annihilation channel is $H\,H\,\rightarrow\,W^+\,W^-$. It is possible to obtain the correct relic density for small mass differences $M_{H^\pm}-M_H\,\lesssim\,10\,\GeV,\,|\lam_{345}|\,\lesssim\,0.25$. 
\end{itemize}

\section{Benchmark Points \label{sec:benchmarks}}

In this section, we list all production cross sections for the production channels
\begin{\eqn}\label{eq:prods}
p\,p\,\rightarrow\,HA,\;HH^+,\;HH^-,\;AH^+,\;AH^-,\;H^+\,H^-,\;AA
\end{\eqn}
for the benchmark scenarios proposed in \cite{Kalinowski:2018ylg}, for center-of-mass energies of 13 and 27 \TeV~ and 100 \TeV~ proton-proton collider. We additionally consider the VBF-like production of $AA$ and $H^+H^-$ at the same hadron collider options as well as a muon-muon collider with center of mass energies of 10\,TeV and 30\,TeV. Cross sections were calculated using Madgraph5 \cite{Alwall:2011uj}, with an UFO input model from \cite{Goudelis:2013uca} (see footnote \ref{foot:ufo}). {We separate the benchmarks} into low mass benchmark points (BPs) {with} dark masses up to 300 \GeV, as well as high mass points (HPs) which cover the whole mass range up to 1 \TeV. The parameter choices as well as kinetic properties of these points are listed in tables \ref{tab:bench} and \ref{tab:bmhigh}. We also emphasize when a point reproduces exact relic density. 

\begin{table}[tbp]
\begin{center}
  \small
  \begin{tabular}{|l|l|l|l|c|c|c|l|l|l|}
\hline
\multirow{2}{*}{No.} & \multirow{2}{*}{$M_H$} & \multirow{2}{*}{$M_A$} & \multirow{2}{*}{$M_{H^\pm}$} & $Z$ & $W$ & DM & \multirow{2}{*}{$\lambda_2$} & \multirow{2}{*}{$\lambda_{345}$} & \multirow{2}{*}{$\Omega_H h^2$}\\[-1mm] 
 & & & & on-shell &  on-shell & $>$50\%  &  &  & \\ \hline
\textbf{BP1} & 72.77 & 107.803 & 114.639 & &&$\checked$& 1.44513 & -0.00440723 & 0.11998\\\hline
BP2 & 65 & 71.525 & 112.85 & &&$\checked$ & 0.779115 & 0.0004 & 0.07076\\ \hline
BP3 & 67.07 & 73.222 & 96.73 &&&$\checked$& 0 & 0.00738 & 0.06159\\ \hline
BP4 & 73.68 & 100.112 & 145.728 & &&$\checked$& 2.08602 & -0.00440723 & 0.089114\\ \hline
\textbf{BP6} & 72.14 & 109.548 & 154.761 & &$\checked$&$\checked$& 0.0125664 & -0.00234 & 0.117\\ \hline
BP7 & 76.55 & 134.563 & 174.367 & &$\checked$&& 1.94779 & 0.0044 & 0.031381\\ \hline
\textbf{BP8} & 70.91 & 148.664 & 175.89 & &$\checked$&$\checked$& 0.439823 & 0.0058 & 0.12207\\ \hline
BP9 & 56.78 & 166.22 & 178.24 & $\checked$&$\checked$&$\checked$& 0.502655 & 0.00338 & 0.081243\\ \hline
BP23 & 62.69 & 162.397 & 190.822 &  $\checked$&$\checked$&$\checked$& 2.63894 & 0.0056 & 0.065\\ \hline
BP10 & 76.69 & 154.579 & 163.045 & &$\checked$& & 3.92071 & 0.0096 & 0.028125\\ \hline
BP11 & 98.88 & 155.037 & 155.438 & &&& 1.18124 & -0.0628 & 0.002735\\ \hline
BP12 & 58.31 & 171.148 & 172.96 &$\checked$&$\checked$&& 0.540354 & 0.00762 & 0.0064104\\ \hline
BP13 & 99.65 & 138.484 & 181.321 & &$\checked$&& 2.46301 & 0.0532 & 0.0012541\\ \hline
\textbf{BP14} & 71.03 & 165.604 & 175.971 & $\checked$&$\checked$&$\checked$& 0.339292 & 0.00596 & 0.11833\\ \hline
\textbf{BP15} & 71.03 & 217.656 & 218.738 & $\checked$&$\checked$&$\checked$& 0.766549 & 0.00214 & 0.12217\\ \hline
\textbf{BP16} & 71.33 & 203.796 & 229.092 & $\checked$&$\checked$&$\checked$& 1.03044 & -0.00122 & 0.12205\\ \hline
BP18 & 147 & 194.647 & 197.403 &&&& 0.387 & -0.018 & 0.0017711\\ \hline
BP19 & 165.8 & 190.082 & 195.999 &&&& 2.7675 & -0.004 & 0.0028308\\ \hline
BP20 & 191.8 & 198.376 & 199.721 & &&& 1.5075 & 0.008 & 0.0084219\\ \hline
\textbf{BP21} & 57.475 & 288.031 & 299.536 & $\checked$&$\checked$&$\checked$& 0.929911 & 0.00192 & 0.11942\\ \hline
\textbf{BP22} & 71.42 & 247.224 & 258.382 &  $\checked$&$\checked$&$\checked$& 1.04301 & -0.0032 & 0.12206\\ \hline
\end{tabular}

\caption{ In all benchmarks $M_h = 125.1$ \GeV. Bold font denotes BP with 100\% DM relic density.
  Note that BP5 and BP17 were excluded by the updated  XENON1T limits \cite{Aprile:2018dbl}. Taken from \cite{Kalinowski:2018ylg}, with adjustments for $\lambda_{345}$ as discussed in \cite{Kalinowski:2018kdn} {and updated relic density values using \texttt{micOMEGAS$\_$5.0.4}}.
  \label{tab:bench}
}
\end{center}
\end{table}

\begin{table}[tbp]
\begin{center}
\small
\begin{tabular}{|l|l|l|l|c|c|c|l|l|l|}
\hline 
\multirow{2}{*}{No.} & \multirow{2}{*}{$M_H$} & \multirow{2}{*}{$M_A$} & \multirow{2}{*}{$M_{H^\pm}$} & $Z$ & $W$ & DM & \multirow{2}{*}{$\lambda_2$} & \multirow{2}{*}{$\lambda_{345}$} & \multirow{2}{*}{$\Omega_H h^2$}\\[-1mm] 
 & & & & on-shell &  on-shell & $>$50\%  &  &  & \\ 
\hline 
HP1 & 176 & 291.36 & 311.96 & $\checked$ & $\checked$ &           &  1.4895 & -0.1035 & 0.00072692\\ \hline 
HP2 & 557 & 562.316 & 565.417 &           &           & $\checked$  &  4.0455 & -0.1385 & 0.07163 \\ \hline 
HP3 & 560 & 616.32 & 633.48 &           &           &           &  3.3795 & -0.0895 & 0.0011357 \\ \hline 
HP4 & 571 & 676.534 & 682.54 & $\checked$ & $\checked$ &           &  1.98 & -0.471 & 0.00056712 \\ \hline 
HP5 & 671 & 688.108 & 688.437 &           &           &           &  1.377 & -0.1455 & 0.024523 \\ \hline 
HP6 & 713 & 716.444 & 723.045 &           &           &           &  2.88 & 0.2885 & 0.035145 \\ \hline 
HP7 & 807 & 813.369 & 818.001 &           &           &           &  3.6675 & 0.299 & 0.032488 \\ \hline 
HP8 & 933 & 939.968 & 943.787 &           &           & $\checked$  &  2.9745 & -0.2435 & 0.09637 \\ \hline 
HP9 & 935 & 986.22 & 987.975 &           &           &           &  2.484 & -0.5795 & 0.0028109 \\ \hline 
\bf HP10 & 990 & 992.36 & 998.12 &           &           & $\checked$  &  3.3345 & -0.040 & 0.12215 \\ \hline 
HP11 & 250.5 & 265.49 & 287.226 &           &           &           &  3.90814 & -0.150071 & 0.0053534 \\ \hline 
HP12 & 286.05 & 294.617 & 332.457 &           &           &           &  3.29239 & 0.112124 & 0.002771 \\ \hline 
HP13 & 336 & 353.264 & 360.568 &           &           &           &  2.48814 & -0.106372 & 0.009366 \\ \hline 
HP14 & 326.55 & 331.938 & 381.773 &           &           &           &  0.0251327 & -0.0626727 & 0.0035646 \\ \hline 
HP15 & 357.6 & 399.998 & 402.568 &           &           &           &  2.06088 & -0.237469 & 0.0034553 \\ \hline 
HP16 & 387.75 & 406.118 & 413.464 &           &           &           &  0.816814 & -0.208336 & 0.01158 \\ \hline 
HP17 & 430.95 & 433.226 & 440.624 &           &           &           &  3.00336 & 0.082991 & 0.032697 \\ \hline 
HP18 & 428.25 & 453.979 & 459.696 &           &           &           &  3.87044 & -0.281168 & 0.0085817 \\ \hline 
HP19 & 467.85 & 488.604 & 492.329 &           &           &           &  4.12177 & -0.252036 & 0.013879 \\ \hline 
HP20 & 505.2 & 516.58 & 543.794 &           &           &           &  2.53841 & -0.354 & 0.0088693 \\ \hline 
\end{tabular}

\caption{High-mass benchmark points (HPs) accessible at colliders with $\mathcal{O}\lb\TeV\rb$ center-of-mass energies. $M_h=125.1\,\GeV$ for all points. HP10 provides exact relic density.  Taken from \cite{Kalinowski:2018ylg}, {with adjustments for $\lambda_{345}$ as discussed in \cite{Kalinowski:2018kdn}} {and updated relic density values using \texttt{micOMEGAS$\_$5.0.4}}. \label{tab:bmhigh} }
\end{center}
\end{table}

Figure \ref{fig:benchmarks} shows the initial benchmark candidates discussed in \cite{Kalinowski:2018ylg}, that obey all current constraints, in the $(M_{H^+}-M_H; M_A-M_H)$ plane. All points form a narrow band corresponding to $M_A \lesssim M_{H^\pm}$. Our chosen benchmark points, also indicated in Fig.~\ref{fig:benchmarks} (red points) cover mass gaps up to about 250\,GeV.

\begin{figure}[tb]
\begin{center}
  \includegraphics[width=0.6\textwidth]{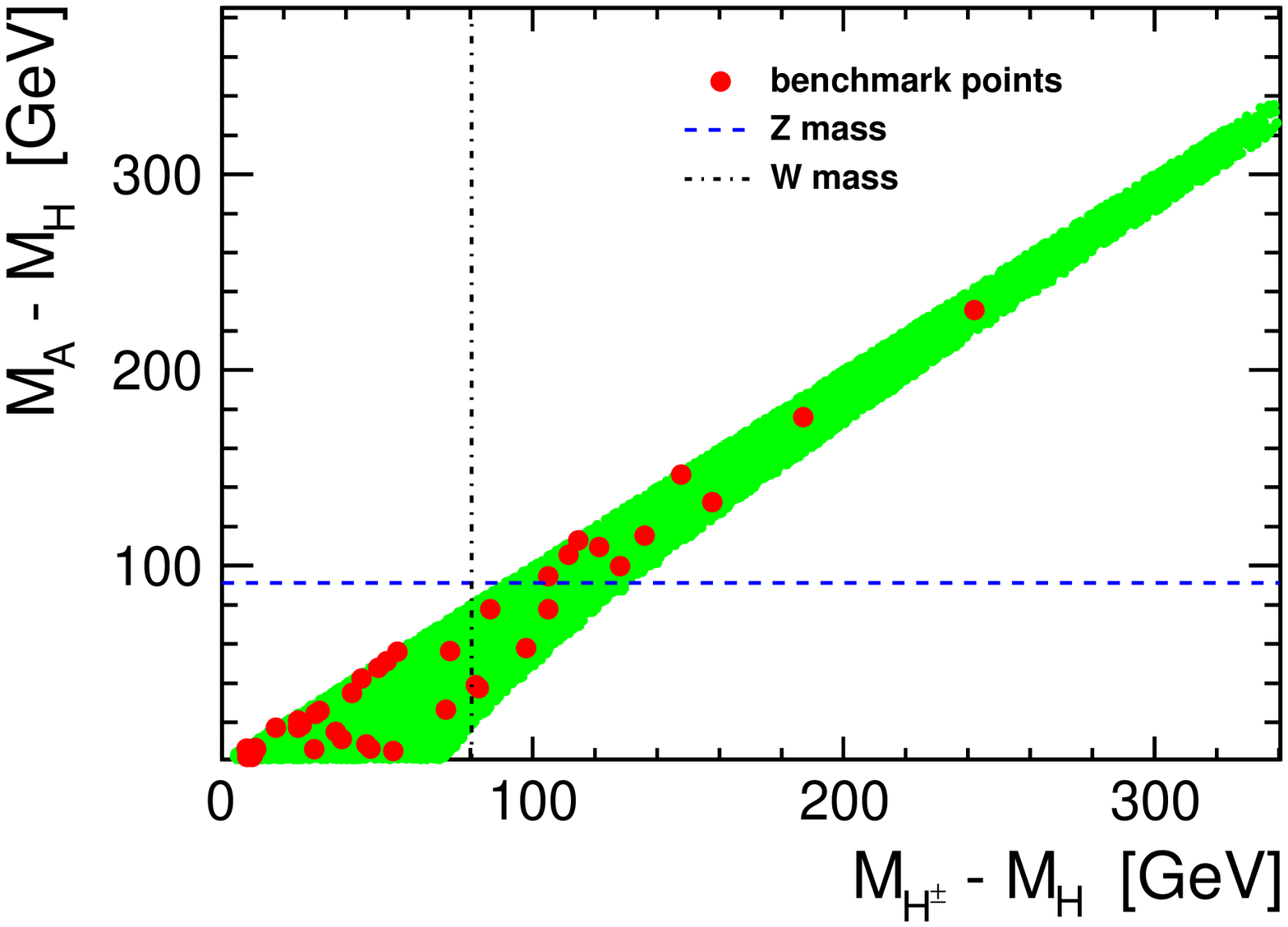}
\end{center}  
\caption{Distribution of benchmark candidate points (green) in the $(M_{H^+}-M_H; M_A-M_H)$ plane{, after all constraints are taken into account}, as well as selected benchmark points (red) in the same plane. The dashed lines indicate the electroweak gauge boson masses that distinguish between on- and off-shell decays of dark scalars. The relatively narrow band stems mainly from electroweak precision constraints. Taken from \cite{Kalinowski:2018ylg}.
\label{fig:benchmarks}}
\end{figure}

\section{Production cross sections at various collider options}\label{sec:sel}

We first focus on {the completed LHC Run 2} with a center-of-mass energy of 13 \TeV~and assuming an integrated luminosity of $150\,\fb^{-1}$. 
The production cross sections at 13\,TeV for all considered benchmark scenarios are listed in tables \ref{tab:xs13low} and \ref{tab:xs13high}.
In Fig.~\ref{fig:cros13} cross sections for different on-shell scalar pair-production channels are compared, shown as a function of the sum of produced scalar masses. We note that, apart from $AA$ production, all processes show a similar decrease in the cross section as the mass scale rises; as these production modes are stemming from Drell-Yan processes with intermediate gauge bosons, the masses remain the only undetermined parameters, while all couplings are given by SM electroweak variables. Therefore, differences between e.g. $H H^+$ and $A H^+$ are small for the same mass scale. In general, $AH^-/H H^-$ states are produced with slightly lower cross sections, due to the parton content of the proton. For the $AA$ process, however, the coupling 
\begin{\eqn}\label{eq:l345b}
\bar{\lam}_{345}\,\equiv\,\lam_3+\lam_4-\lam_5\,=\,\lam_{345}-2\,\frac{M_H^2-M_A^2}{v^2}
\end{\eqn}
 determines the cross section, which is no longer a function of the mass only. Therefore, for this production mode the cross sections do not follow the same simple behaviour. For example, cross sections $\lesssim\,0.1\fb$ are usually achieved for $\bar{\lam}_{345}\,\lesssim\,0.5$ for the low mass BPs.

\begin{table}[p]
\small
\begin{center}
\begin{tabular}{|l|l|l|l|c|c|c|c|c|c|c|}
\hline 
No. & $M_H$ & $M_A$ & $M_{H^\pm}$ & $HA $  & $ H\,H^+ $ & $H\,H^-$ &  $A H^+ $&$A\,H^-$  & $ H^+H^- $ & $AA$ \\ \hline 
\bf BP1 & 72.77 & 107.803 & 114.639 & 322 &304&183&169&98.2&133&0.925 \\ \hline 
BP2 & 65 & 71.525 & 112.85 & 1020 & 363 &220& 323 &195& 141 & 1.46 \\ \hline 
BP3 & 67.07 & 73.222 & 96.73 & 909 & 505 &311& 444 &272& 243 & 0.939\\ \hline 
BP4 & 73.68 & 100.112 & 145.728 &377 &166 &96.4& 115 &65.7 & 56.3 &0.757 \\ \hline 
\bf BP6 & 72.14 & 109.548 & 154.761 &314 & 144 &83.5& 90.0 & 50.0& 45.5 &0.912 \\ \hline 
BP7 & 76.55 & 134.563 & 174.367 & 173 & 99.1 &56.2& 50.9 &27.7& 29.3 &0.491 \\ \hline 
\bf BP8 & 70.91 & 148.664 & 175.89 & 144 & 103 &58.6& 42.8 &23.0& 28.6 &0.500 \\ \hline 
BP9 & 56.78 & 166.22 & 178.24 & 125 & 116 &66.4& 34.5 &18.3& 27.3 & 0.683\\ \hline 
BP10 & 76.69 & 154.579 & 163.045 & 120 & 119 &67.8& 46.4 &25.2& 37.3 &0.489 \\ \hline 
BP11 & 98.88 & 155.037 & 155.438 & 87.7 & 101 &57.2& 50.5 &27.5& 44.0 & 0.278\\ \hline 
BP12 & 58.31 & 171.148 & 172.96 & 113 & 125 &71.7& 34.6 &18.4& 30.3 & 0.554\\ \hline 
BP13 & 99.65 & 138.484 & 181.321 & 113 & 68.8 &38.2& 44.9 &24.2 & 25.0 &0.209 \\ \hline 
\bf BP14 & 71.03 & 165.604 & 175.971 & 106 & 103 &58.5& 35.6 &18.9 &28.6 &0.650 \\ \hline 
\bf BP15 & 71.03 & 217.656 & 218.738 & 46.9 & 54.6 &30.0& 14.2 &7.14 & 12.9 &0.502 \\ \hline 
\bf BP16 & 71.33 & 203.796 & 229.092 & 57.3 & 47.3 &25.8& 14.6 &7.36& 10.9 & 0.536\\ \hline 
BP18 & 147 & 194.647 & 197.403 & 29.2 & 34.0 &18.1& 21.3 &11.0& 17.9 &0.112 \\ \hline 
BP19 & 165.8 & 190.082 & 195.999 & 25.2 & 28.6 &15.0& 22.6 &11.7& 18.3 &0.0362 \\ \hline 
BP20 & 191.8 & 198.376 & 199.721 & 17.7 & 21.4 &11.0& 20.1 &10.3& 16.9 & 0.00305 \\ \hline 
\bf BP21 & 57.475 & 288.031 & 299.536 & 20.6 & 21.8 &11.4& 4.44 &2.06 &4.09&0.345 \\ \hline 
\bf BP22 & 71.42 & 247.224 & 258.382 & 31.3 & 32.5 &17.3& 8.04 &3.89 &7.00 &0.381\\ \hline 
BP23 & 62.69 & 162.397 & 190.822 & 125 & 88.9 &50.2& 31.3 &16.5 & 21.1 & 0.545 \\ \hline 
\end{tabular}

\caption{ \label{tab:xs13low} Production cross sections in \fb~ for low-mass benchmark points from table \ref{tab:bench}, for different on-shell scalar pair-production channels at 13 \TeV~LHC. {Bold font denotes benchmark points for which H completely saturates DM relic density.}  }
\end{center}
\end{table}

\begin{table}[p]
\begin{center}
\small
\begin{tabular}{|l|l|l|l|c|c|c|c|c|c|c|}
\hline 
No. & $M_H$ & $M_A$ & $M_{H^\pm}$ & $HA $  & $ H\,H^+ $ & $H\,H^-$ &  $A H^+ $ &$ A H^-$ & $ H^+H^- $ & $AA$ \\ \hline 
 
HP1 & 176 & 291.36 & 311.96 & 8.33 & 8.76 &4.27& 3.99 &1.84&  3.12 & 0.132\\ \hline
HP2 & 557 & 562.316 & 565.417 & 0.184 & 0.259 &0.0993& 0.253 &0.0970 &  0.190 & -\\ \hline
HP3 & 560 & 616.32 & 633.48 & 0.143 & 0.191 &0.0718& 0.153  &0.0565&  0.115 & 0.00273 \\ \hline
HP4 & 571 & 676.534 & 682.54 & 0.105 & 0.149 &0.0552& 0.0991 &0.0358 &  0.0830 & 0.00512\\ \hline
HP5 & 671 & 688.108 & 688.437 & 0.0672 & 0.0990 &0.0358& 0.0927  &0.0334&  0.0690 & - \\ \hline
HP6 & 713 & 716.444 & 723.045 & 0.0511 & 0.0740 &0.0263& 0.0730  &0.0260&  0.0529 & - \\ \hline
HP7 & 807 & 813.369 & 818.001 & 0.0253 & 0.0375 &0.0129& 0.0367  &0.0126&  0.0265 & -\\ \hline
HP8 & 933 & 939.968 & 943.787 & 0.0106 & 0.0161 &0.00530& 0.0157 &0.00518&  0.0113 & -\\ \hline
HP9 & 935 & 986.22 & 987.975 & 0.00904 & 0.0139 &0.00453& 0.0118  &0.00383&  0.00883 & -\\ \hline
\bf HP10 & 990 & 992.36 & 998.12 & 0.00742 & 0.0113 &0.00366& 0.0112  &0.00363&  0.00794 & -\\ \hline
HP11 & 250.5 & 265.49 & 287.226 & 5.82 & 6.30 &3.00& 5.66  &2.68&  4.03 & - \\ \hline
HP12 & 286.05 & 294.617 & 332.457 & 3.59 & 3.60 &1.64& 3.41  &1.56&  2.23 & 0.00337 \\ \hline
HP13 & 336 & 353.264 & 360.568 & 1.73 & 2.21 &0.977& 1.99  &0.874&  1.54 & 0.00135 \\ \hline
HP14 & 326.55 & 331.938 & 381.773 & 2.11 & 2.05 &0.902& 1.99  &0.872&  1.23 & -\\ \hline
HP15 & 357.6 & 399.998 & 402.568 & 1.14 & 1.52 &0.655& 1.21  &0.512&  0.955 & 0.00556 \\ \hline
HP16 & 387.75 & 406.118 & 413.464 & 0.931 & 1.21 &0.515& 1.10  &0.464&  0.840 & - \\ \hline
HP17 & 430.95 & 433.226 & 440.624 & 0.632 & 0.837 &0.347& 0.828 &0.342&  0.627 & -\\ \hline
HP18 & 428.25 & 453.979 & 459.696 & 0.575 & 0.769 &0.318& 0.678  &0.276&  0.517 &- \\ \hline
HP19 & 467.85 & 488.604 & 492.329 & 0.394 & 0.541 &0.217& 0.490  &0.196&  0.374 & - \\ \hline
HP20 & 505.2 & 516.58 & 543.794 & 0.287 & 0.357 &0.140& 0.340  &0.132&  0.233 & - \\ \hline
\end{tabular}

\caption{\label{tab:xs13high} Production cross sections in \fb~ for high-mass benchmark points from table \ref{tab:bmhigh}, for different on-shell scalar pair-production channels at the 13 \TeV~LHC. Dashes ( - ) indicate cross-section values smaller than $10^{-3}\,\fb$. {For the HP10 scenario (bold) H completely saturates DM relic density.} }
\end{center}
\end{table}

\begin{figure}[tb]
\begin{center}
  \includegraphics[width=0.6\textwidth]{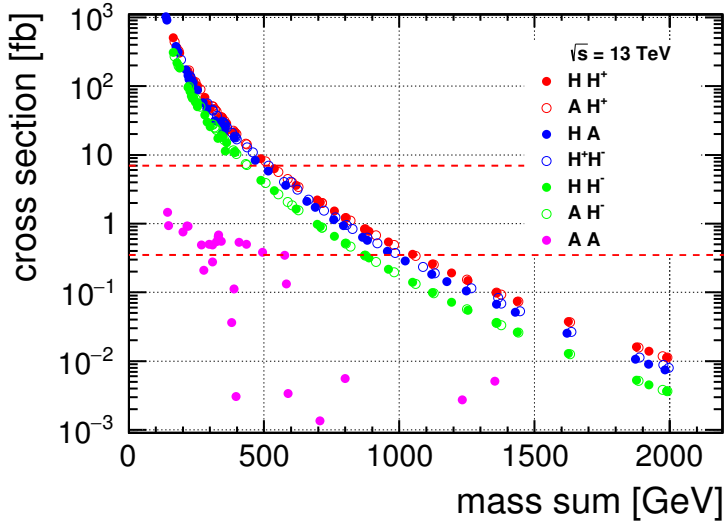}
\end{center}  
\caption{ Production cross sections for benchmarks from tables \ref{tab:bench} and \ref{tab:bmhigh} as a function of the produced scalar mass sum, for on-shell scalar pair-production at 13\,TeV LHC. Horizontal dashed lines indicate minimal cross sections required to produce 1000 events at LHC Run 2 and HL-LHC (see text for details).
\label{fig:cros13}}
\end{figure}

We label scenarios as realistic,\footnote{We note that this only corresponds to a rough estimate of accessibility. Detailed studies, including background simulation, would be needed to determine discovery options for each BP individually.} if they produce at least 1000 events during that run, translating to minimal cross sections of about 7 \fb { indicated by the horizontal dashed line in Fig.~\ref{fig:cros13}.} Note that the decays {of the heavier dark scalars} are predominantly given by 
\begin{\eqn*}
H^\pm\,\rightarrow\,W^\pm\,H;\,A\,\rightarrow\,Z\,H
\end{\eqn*}
with the electroweak gauge bosons decaying as in the SM. Only BPs 2,3,4 have sizeable branching ratios for the channel $H^\pm\,\rightarrow\,A\,W^\pm$ of 0.34, 0.25, and 0.08 respectively.

\subsection{Current LHC data, Run 2}

With the simple counting criterium proposed above, one can see that minimum cross section of 7\,fb (horizontal dashed line in  Fig.~\ref{fig:cros13}) limits the LHC Run 2 sensitivity to the scalar mass sum of about 450 GeV for $HH^-$ and $AH^-$ production channels and to about 500 GeV for other scalar pair-production channels. We see that most of the low mass benchmark points in table \ref{tab:xs13low} (BPs 1-16, 18-20 as well as 23) provide high enough cross sections for dark scalar pair-production in all channels but the $AA$ pair-production channel.
On the other hand, for the high-mass benchmark points (table \ref{tab:xs13high}), only HP 1 renders high enough cross sections in the $HA$ and $HH^+$ production mode.

\subsection{High luminosity option}

At the high luminosity LHC, the target integrated luminosity corresponds to $3\, \ab^{-1}$ (see e.g. \cite{hltwiki}), lowering the cross-section threshold for our simple counting criterium to 0.33\,\fb.
The accessible mass range for pair-production of IDM scalars is extended to a mass sum of about 850\,GeV for $HH^-$ and $AH^-$ channel, and about 1\,TeV for other channels (except for $AA$), see Fig.~\ref{fig:cros13}.
The $AA$ channel additionally opens up for BPs 1-10, 12, 14-17, and 23. Similarly BP21 and 22 also render the minimal number of generated events in all channels. Only for BPs 11,13, and 18-20 the total number of events generated does not suffice in the $AA$ channel\footnote{Note that the $hAA$ coupling scales with $\bar{\lam}_{345}$, cf. eqn (\ref{eq:l345b}).}.  For the high-mass points, now HP1, HP11-16 become accessible in all but the $AA$ channel; for HPs 17-19, the $HA,H,\,H\,H^+,\,A\,H^+,\, H^+\,H^-$ channels seem to become accessible, corresponding to a mass range for scalar masses up to 500 \GeV.

\subsection{High energy option}

Values for the production cross sections at a 27 \TeV~ center-of-mass energy are given in tables \ref{tab:xs27low} and \ref{tab:xs27high}.
With a center-of-mass energy of 27 \TeV~and a target luminosity of 15 $\ab^{-1}$ \cite{Abada:2019ono}, the minimal cross section required to obtain at least 1000 events in the full run further decreases to  0.07 \fb. This means that all but BP20 are accessible in all channels. BP20 features a low value of $\bar{\lam}_{345}\,\sim\,0.09$ and a relatively high mass $M_A$, leading to a low $AA$ cross section even at the 27 \TeV~ HE-LHC. For the high-mass points, HPs 2-7, 11-20 are open in all but the $AA$ channel, while HP1 even renders a large enough cross section for this channel as well. For HPs 8-10, the $H\,H^-,\,A\,H^-$ channels additionally remain inaccessible. This means that all HPs and BPs are accessible in at least one channel, with scalar masses up to 1 \TeV. The enhancement factors for production processes with respect to cross sections at the LHC including at least one unstable new scalar are shown in Fig.~\ref{fig:ratios}. In  general, for the low BPs the cross-section enhancement is about a factor 3, where for $AA$ final states a maximal value of $\sim\,6$ is reached for BP21. For HPs the enhancement can be up to a factor 10 depending on the dark scalar masses\footnote{In fact, the largest enhancement is obtained for HP10, where the cross section increases by a factor 20. However, the absolute value for $AA$ production at 27 \TeV for this point is $\mathcal{O}\lb 10^{-6}\,\fb \rb$, making it too small for a detailed investigation of this channel.}. 
\begin{table}[p]
\small
\begin{center}
\begin{tabular}{|l|l|l|l|c|c|c|c|c|c|c|}
\hline 
No. & $M_H$ & $M_A$ & $M_{H^\pm}$ & $HA $  & $ H\,H^+ $ & $H\,H^-$ &  $A H^+ $  &$A\,H^-$ & $ H^+H^- $ & $AA$ \\ \hline 
\bf BP1 & 72.77 & 107.803 & 114.639 & 846 &770&516 &441&289&368&3.84 \\ \hline 
BP2 & 65 & 71.525 & 112.85 & 2540 & 910 &614& 814 &547& 387 & 5.23 \\ \hline 
BP3 & 67.07 & 73.222 & 96.73 & 2270 & 1250 &849& 1100 &750& 645 & 3.39\\ \hline 
BP4 & 73.68 & 100.112 & 145.728 &982 &432 &284& 308 &199& 166 &3.06 \\ \hline 
\bf BP6 & 72.14 & 109.548 & 154.761 &824 & 380 &248& 241 &154& 137 &3.81 \\ \hline 
BP7 & 76.55 & 134.563 & 174.367 & 470 & 266 &172& 143 &89.4 & 91.7 & 2.22 \\ \hline 
\bf BP8 & 70.91 & 148.664 & 175.89 & 396 & 276 &178& 122 &75.5& 90.5 &2.34 \\ \hline 
BP9 & 56.78 & 166.22 & 178.24 & 347 & 309 &200& 100 &61.2& 87.0 & 3.35\\ \hline 
BP10 & 76.69 & 154.579 & 163.045 & 332 & 316 &205&131 & 81.9& 114 &2.33 \\ \hline 
BP11 & 98.88 & 155.037 & 155.438 & 249 & 271 &175& 142 &88.8& 131 & 1.33\\ \hline 
BP12 & 58.31 & 171.148 & 172.96 & 313 & 331 &215& 100 &61.5& 94.7 & 2.76\\ \hline 
BP13 & 99.65 & 138.484 & 181.321 & 316 & 189 &120& 127 &79.0& 79.1 &0.954 \\ \hline 
\bf BP14 & 71.03 & 165.604 & 175.971 & 297 & 276 &178& 103 &63.1& 90.3 &3.19 \\ \hline 
\bf BP15 & 71.03 & 217.656 & 218.738 & 138 &152 &95.6& 44.3 &26.0& 45.0 &2.78 \\ \hline 
\bf BP16 & 71.33 & 203.796 & 229.092 & 167 & 133 &83.2& 45.4 & 26.8& 38.9 & 2.87\\ \hline 
BP18 & 147 & 194.647 & 197.403 & 89.6 & 98.5 &60.4& 64.0 &38.4 & 57.8 &0.590 \\ \hline 
BP19 & 165.8 & 190.082 & 195.999 & 78.4 & 83.9 &51.1& 67.5 &40.6 & 58.5 &0.188 \\ \hline 
BP20 & 191.8 & 198.376 & 199.721 & 56.7 & 64.3 &38.6& 60.7 & 36.4& 54.4 & 0.0161 \\ \hline 
\bf BP21 & 57.475 & 288.031 & 299.536 & 64.6 & 65.0 &39.2& 15.5 &8.62 &17.5 &2.21 \\ \hline 
\bf BP22 & 71.42 & 247.224 & 258.382 & 94.9 & 94.0 &57.8& 26.4 &15.1&26.9 &2.25\\ \hline 
BP23 & 62.69 & 162.397 & 190.822 & 348 & 241 &154& 91.2 &55.9& 69.3 & 2.66 \\ \hline 
\end{tabular}

\caption{ \label{tab:xs27low} Production cross sections for BPs from table \ref{tab:bench} in \fb~for for on-shell scalar pair-production at 27 \TeV~HE-LHC. }
\end{center}
\end{table}

\begin{table}[p]
\begin{center}
\small
\begin{tabular}{|l|l|l|l|c|c|c|c|c|c|c|}
\hline 
No. & $M_H$ & $M_A$ & $M_{H^\pm}$ & $HA $  & $ H\,H^+ $ &  $H H^-$ &  $A H^+ $  &$A H^-$ &$ H^+H^- $ & $AA$ \\ \hline 
 
HP1 & 176 & 291.36 & 311.96 & 28.7 & 28.5 &16.4& 14.1  &7.79&  12.9 & 0.850\\ \hline
HP2 & 557 & 562.316 & 565.417 & 1.07 & 1.36 &0.650& 1.34  &0.637&  1.11 & -\\ \hline
HP3 & 560 & 616.32 & 633.48 & 0.871 & 1.06 &0.498& 0.886  &0.410&  0.787 & 0.0311 \\ \hline
HP4 & 571 & 676.534 & 682.54 & 0.678 & 0.871 &0.402& 0.630  &0.284&  0.642 & 0.0644\\ \hline
HP5 & 671 & 688.108 & 688.437 & 0.474 & 0.629 &0.284& 0.598  &0.269&  0.488 & 0.00151 \\ \hline
HP6 & 713 & 716.444 & 723.045 & 0.381 & 0.499 &0.222& 0.494  &0.220&  0.395 & - \\ \hline
HP7 & 807 & 813.369 & 818.001 & 0.219 & 0.294 &0.126& 0.289  &0.123&  0.230 & -\\ \hline
HP8 & 933 & 939.968 & 943.787 & 0.113 & 0.155 &0.0634& 0.152 &0.0623&  0.118 & -\\ \hline
HP9 & 935 & 986.22 & 987.975 & 0.0999 & 0.138 &0.0563& 0.123  &0.0496&  0.103 & 0.00364\\ \hline
\bf HP10 & 990 & 992.36 & 998.12 & 0.0861 & 0.119 &0.0479& 0.118  &0.0476&  0.0910 & -\\ \hline
HP11 & 250.5 & 265.49 & 287.226 & 20.9 & 21.2 &12.0& 19.3  &10.8& 15.1 & 0.00521 \\ \hline
HP12 & 286.05 & 294.617 & 332.457 & 13.6 & 12.9 &7.07& 12.3  &6.72&  9.09 & 0.0219 \\ \hline
HP13 & 336 & 353.264 & 360.568 &  7.18 & 8.39 &4.49&  7.67 &4.08& 6.49& 0.00980 \\ \hline
HP14 & 326.55 & 331.938 & 381.773 & 8.53 & 7.86 &4.19& 7.65  &4.07&  5.43 & -\\ \hline
HP15 & 357.6 & 399.998 & 402.568 & 5.00 & 6.07 &3.19& 4.97  &2.57&  4.35 & 0.0440 \\ \hline
HP16 & 387.75 & 406.118 & 413.464 & 4.20 & 5.00 &2.59& 4.60  &2.37&  3.85 & 0.00448 \\ \hline
HP17 & 430.95 & 433.226 & 440.624 & 3.02 & 3.64 &1.85& 3.61 &1.83&  3.01 & -\\ \hline
HP18 & 428.25 & 453.979 & 459.696 & 2.78 & 3.39 &1.72& 3.05  &1.53&  2.57 &0.00756 \\ \hline
HP19 & 467.85 & 488.604 & 492.329 & 2.02 & 2.52 &1.25& 2.32  &1.14&  1.95 & 0.00385 \\ \hline
HP20 & 505.2 & 516.58 & 543.794 & 1.55 & 1.78 &0.862& 1.71  &0.824&  1.32 & - \\ \hline
\end{tabular}

\caption{\label{tab:xs27high} Production cross sections for HPs from table \ref{tab:bmhigh} in \fb~for high-mass benchmark points for scalar pair-production at the 27 \TeV~HE-LHC. }
\end{center}
\end{table}

\begin{figure}[tb]
\begin{center}
  \includegraphics[width=0.6\textwidth]{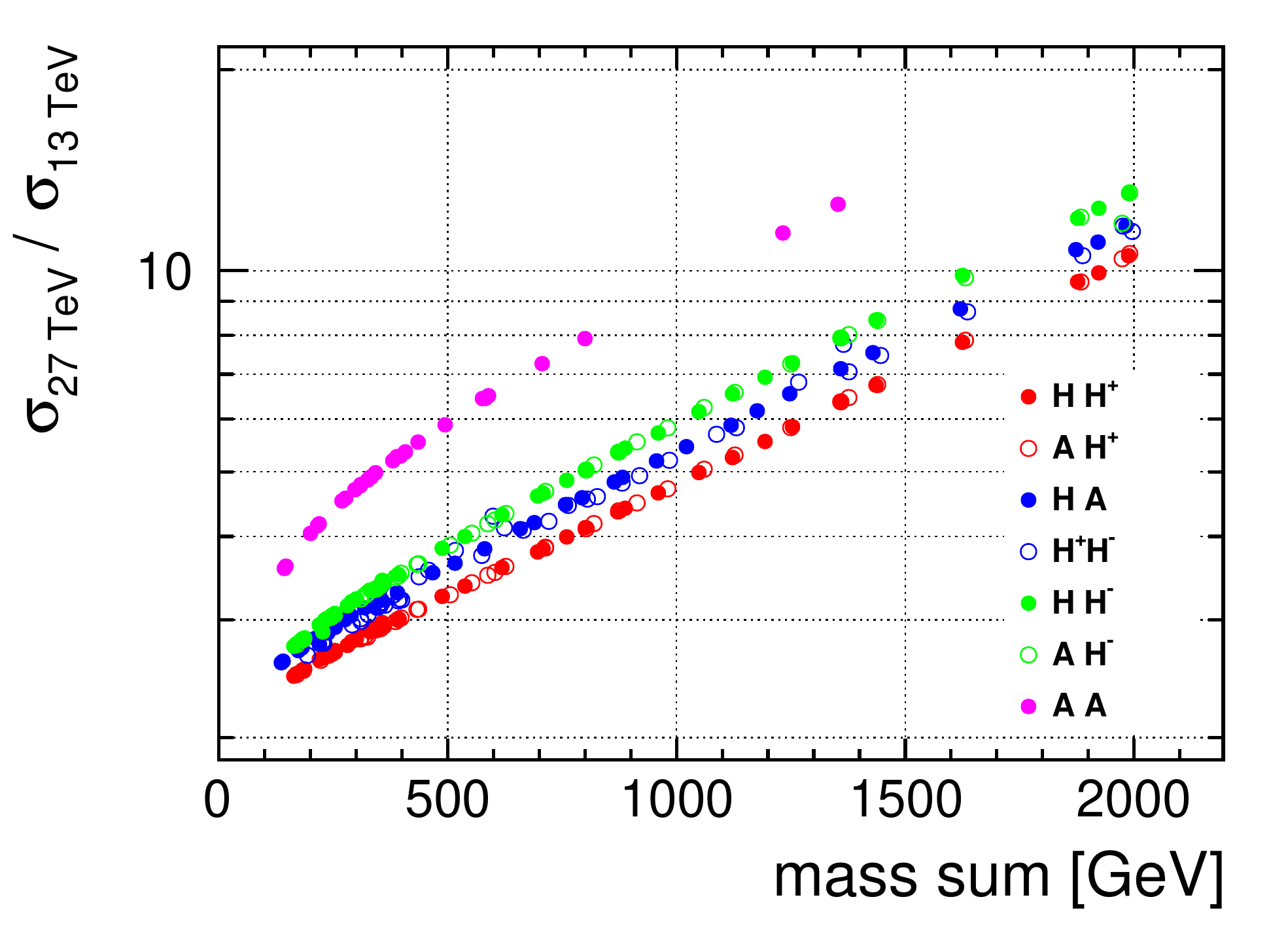}
\end{center}  
\caption{Ratio of production cross sections for all production channels specified with at least one unstable new scalar at the 27 \TeV~HE-LHC and current center-of-mass energy of 13 \TeV. While in the low energy range, cross sections are enhanced roughly by a factor $\lesssim\,3$, for higher masses they can change by an order of magnitude. Scenarios with cross-section value smaller than $10^{-3}\,\fb$ at 13\,TeV are not indicated.
\label{fig:ratios}}
\end{figure}

\subsection{100 \TeV~ proton-proton collider}

A circular hadron-hadron collider with a 100 \TeV~ center of mass energy is currently another option for a future accelerator design \cite{Benedikt:2018csr,Strategy:2019vxc}. For reference, we therefore list the corresponding cross-sections for scalar pair-production in tables \ref{tab:xs100low} and \ref{tab:xs100high}. The target accelerator luminosity corresponds to $20\,\ab^{-1}$; this corresponds to a production cross section of $5\,\times\,10^{-2}\,\fb,$ respectively, to fulfill our accessibility criterium.

\begin{figure}[tb]
\begin{center}
  \includegraphics[width=0.6\textwidth]{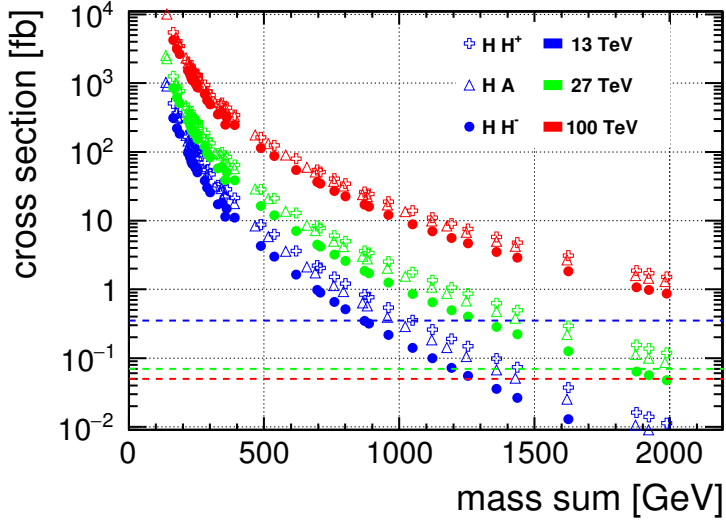}
\end{center}  
\caption{ Production cross sections for benchmarks from tables \ref{tab:bench} and \ref{tab:bmhigh} as a function of the produced scalar mass sum, for selected scalar pair-production channels, for 13\,TeV, 27\,TeV and 100\,TeV proton collider options.
Horizontal dashed lines indicate minimal cross sections required to produce 1000 events at the respective energy, assuming design luminosity.
\label{fig:crosall}}
\end{figure}

For the low BPs, this would allow to close the remaining $AA$ channel for BP 20. For the high-mass benchmark points, HPs 1,3,4,9, 11-13, 15,16,18,19 now could be reachable in all channels using our criterium. For the remaining points, the $AA$ production cross section remains too low. As for the HE-LHC, this corresponds to a possible mass reach up to 1 \TeV~ for the single scalar masses, where however a larger number of total channels is open.
\begin{table}[p]
\small
\begin{center}
\begin{tabular}{|l|l|l|l|c|c|c|c|c|c|c|}
\hline 
No. & $M_H$ & $M_A$ & $M_{H^\pm}$ & $HA $  & $ H\,H^+ $ & $H\,H^-$ &  $A H^+ $&$A\,H^-$  & $ H^+H^- $ & $AA$ \\ \hline 
\bf BP1 & 72.77 & 107.803 & 114.639 &4.00&3.47&2.65&2.06&1.55&1.85& 0.0337\\ \hline
BP2 & 65 & 71.525 & 112.85 &11.2&4.07&3.12&3.67&2.80&1.94&0.0380 \\ \hline
BP3 & 67.07 & 73.222 & 96.73 &10.1&5.47&4.22&4.88&3.75&3.09&0.0249 \\ \hline
BP4 & 73.68 & 100.112 & 145.728 &4.61&2.02&1.52&1.47&1.10&0.901&0.0260 \\ \hline
\bf BP6 & 72.14 & 109.548 & 154.761 &3.91&1.79&1.34&1.17&0.871&0.763&0.0336 \\ \hline
BP7 & 76.55 & 134.563 & 174.367 &2.31&1.29&0.957&0.722&0.529&0.530&0.0215 \\ \hline
\bf BP8 & 70.91 & 148.664 & 175.89 &1.97&1.33&0.992&0.622&0.453&0.533&0.0238 \\ \hline
BP9 & 56.78 & 166.22 & 178.24 &1.74&1.47&1.10&0.517&0.375&0.517&0.0359 \\ \hline
BP10 & 76.69 & 154.579 & 163.045 &1.67&1.51&1.13&0.668&0.488&0.641&0.0241 \\ \hline
BP11 & 98.88 & 155.037 & 155.438 &1.28&1.31&0.975&0.718&0.525&0.715&0.0137 \\ \hline
BP12 & 58.31 & 171.148 & 172.96 &1.58&1.57&1.18&0.519&0.376&0.550&0.0299 \\ \hline
BP13 & 99.65 & 138.484 & 181.321 &1.60&0.937&0.691&0.647&0.472&0.459&0.00938 \\ \hline
\bf BP14 & 71.03 & 165.604 & 175.971 &1.51&1.33&0.989&0.531&0.385&0.532&0.0341 \\ \hline
\bf BP15 & 71.03 & 217.656 & 218.738 &0.742&0.763&0.560&0.244&0.173&0.301&0.0341 \\ \hline
\bf BP16 & 71.33 & 203.796 & 229.092 &0.882&0.674&0.493&0.250&0.177&0.268&0.0341 \\ \hline
BP18 & 147 & 194.647 & 197.403 &0.499&0.511&0.370&0.343&0.246&0.337&0.00685 \\ \hline
BP19 & 165.8 & 190.082 & 195.999 &0.441&0.441&0.318&0.361&0.259&0.336&0.00216 \\ \hline
BP20 & 191.8 & 198.376 & 199.721 &0.329&0.345&0.247&0.327&0.234&0.311&0.000189 \\ \hline
\bf BP21 & 57.475 & 288.031 & 299.536 &0.367&0.346&0.249&0.0941&0.0646&0.153&0.0319 \\ \hline
\bf BP22 & 71.42 & 247.224 & 258.382 &0.524&0.487&0.353&0.152&0.106&0.204&0.0296\\ \hline
BP23 & 62.69 & 162.397 & 190.822 &1.74&1.17&0.867&0.476&0.345&0.425&0.0280 \\ \hline
\end{tabular}

\caption{ \label{tab:xs100low} Production cross sections for BPs from table \ref{tab:bench} in \pb~ for for on-shell scalar pair-production at a 100 \TeV~ FCC. }
\end{center}
\end{table}

\begin{table}[p]
\begin{center}
\small
\begin{tabular}{|l|l|l|l|c|c|c|c|c|c|c|}
\hline 
No. & $M_H$ & $M_A$ & $M_{H^\pm}$ & $HA $  & $ H\,H^+ $ & $H\,H^-$ &  $A H^+ $ & $A\,H^-$ & $ H^+H^- $ & $AA$ \\ \hline 
 
HP1 & 176 & 291.36 & 311.96 &176&163&114&86.4&59.2&103&12.3 \\ \hline
HP2 & 557 & 562.316 & 565.417  &9.89&11.1&7.00&10.9&6.88&10.1&- \\ \hline
HP3 & 560 & 616.32 & 633.48  &8.32&9.01&5.62&7.72&4.77&9.27&0.781 \\ \hline
HP4 & 571 & 676.534 & 682.54  &6.76&7.60&4.70&5.79&3.53&9.13&1.76 \\ \hline
HP5 & 671 & 688.108 & 688.437  &5.02&5.78&3.52&5.54&3.37&5.19&0.0421 \\ \hline
HP6 & 713 & 716.444 & 723.045  &4.21&4.79&2.89&4.75&2.86&4.39 &0.0185 \\ \hline 
HP7 & 807 & 813.369 & 818.001  &2.69&3.11&1.84&3.07&1.81&2.85 &0.0210 \\ \hline
HP8 & 933 & 939.968 & 943.787  &1.59&1.87&1.07&1.85&1.06&1.66&- \\ \hline
HP9 & 935 & 986.22 & 987.975  &1.45&1.71&0.978&1.56&0.886&1.72 & 0.150\\ \hline
\bf HP10 & 990 & 992.36 & 998.12  &1.29&1.52&0.863&1.52&0.856&1.36 &- \\ \hline
HP11 & 250.5 & 265.49 & 287.226  &132&125&86.6&115&79.2&99.1 &0.0714 \\ \hline
HP12 & 286.05 & 294.617 & 332.457  &89.9&79.5&54.3&76.1&51.9&65.6 &0.320 \\ \hline
HP13 & 336 & 353.264 & 360.568  &51.0&54.2&36.4&50.1&33.6&46.7 & 0.160\\ \hline
HP14 & 326.55 & 331.938 & 381.773  &59.4&51.2&34.3&49.9&33.5&42.2 &0.00751 \\ \hline
HP15 & 357.6 & 399.998 & 402.568  &37.2&40.7&27.0&34.1&22.5&33.7 &0.781 \\ \hline
HP16 & 387.75 & 406.118 & 413.464  &31.8&34.3&22.6&31.8&20.9&29.6 &0.0805 \\ \hline
HP17 & 430.95 & 433.226 & 440.624  &23.9&26.0&17.0&25.7&16.8&23.7 & 0.0180\\ \hline
HP18 & 428.25 & 453.979 & 459.696  &22.3&24.4&15.9&22.2&14.4&20.9 &0.147 \\ \hline
HP19 & 467.85 & 488.604 & 492.329  &16.9&18.8&12.1&17.5&11.2&16.5 &0.0795 \\ \hline
HP20 & 505.2 & 516.58 & 543.794  &13.5&14.0&8.87&13.5&8.54&12.1&- \\ \hline
\end{tabular}

\caption{\label{tab:xs100high} Production cross sections for HPs from table \ref{tab:bmhigh} in \fb~for high-mass benchmark points for scalar pair-production at a 100 \TeV~FCC.  }
\end{center}
\end{table}

Production cross sections for selected scalar pair-production channels, for different  proton collider options, are compared in Fig.~\ref{fig:crosall}.
In general, production cross sections are enhanced by one to two orders of magnitude with respect to the corresponding values at the 13 \TeV LHC, cf. Fig \ref{fig:ratios_100}.
\begin{figure}[tb]
\begin{center}
  \includegraphics[width=0.6\textwidth]{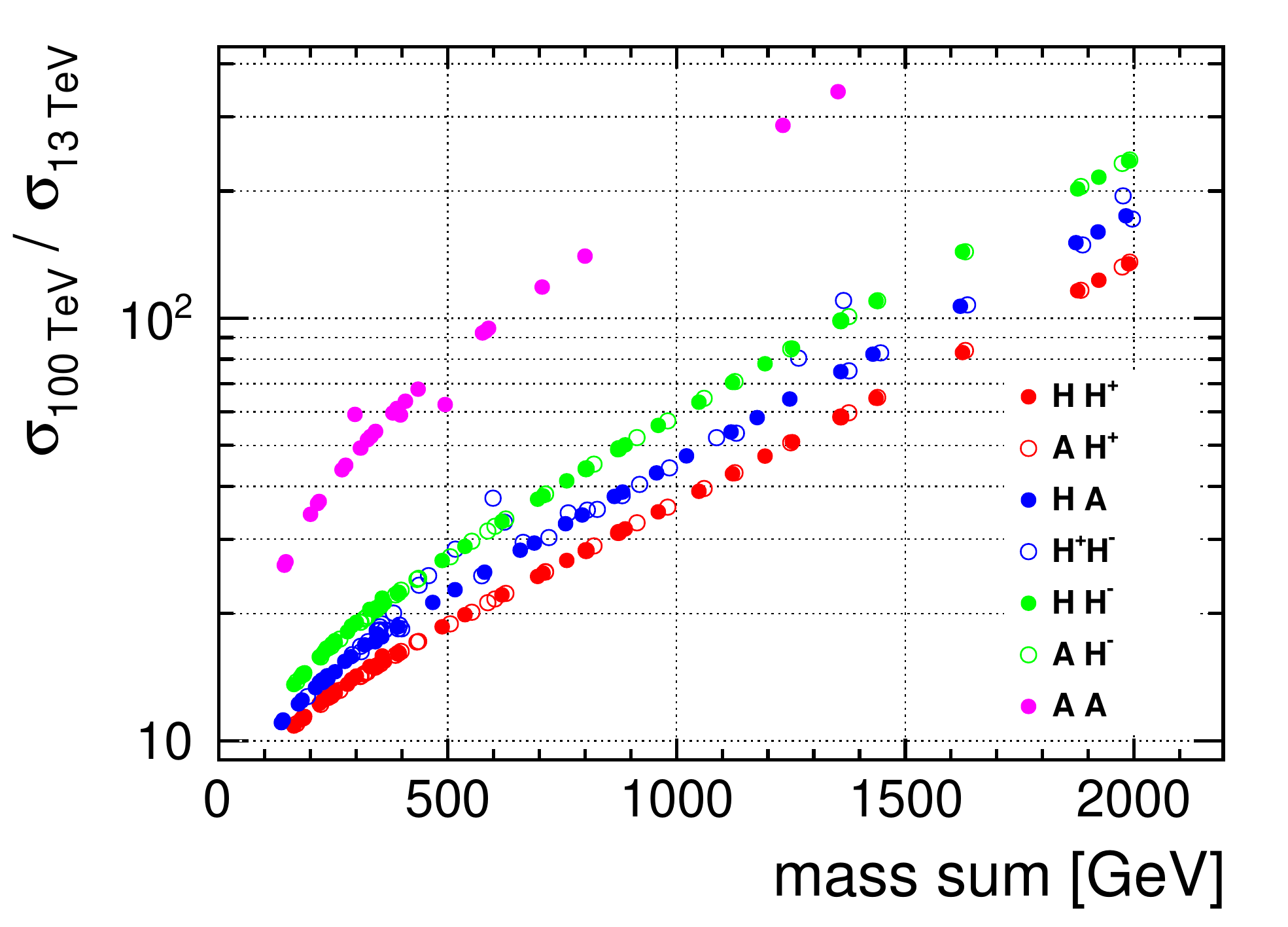}
\end{center}  
\caption{Ratio of production cross sections for all production channels specified with at least one unstable new scalar at the 100 \TeV~ pp collider and the LHC with a center-of-mass energy of 13 \TeV. Production cross sections are enhanced roughly by one to two orders of magnitude. 
\label{fig:ratios_100}}
\end{figure}

\subsection{VBF-like topologies}

Apart from the direct pair-production processes in Eqn. (\ref{eq:prods}), also final states with additional jets should be considered. We here include all processes that lead to the required final state; a subset of these are VBF-like topologies. As an example, we additionally consider
\begin{\eqn*}
p\,p\,\rightarrow\,A\,A\,j\,j,\; p\,p\,\rightarrow\,H^+\,H^-\,j\,j.
\end{\eqn*}
Both processes can include VBF-type diagrams. The respective cross sections for the low and high mass benchmarks, with varying collider energies, are given in tables \ref{tab:VBF_low} and \ref{tab:VBF_high}, and compared in Fig.~\ref{fig:crosvbf}.
\begin{figure}[tb]
\begin{center}
  \includegraphics[width=0.6\textwidth]{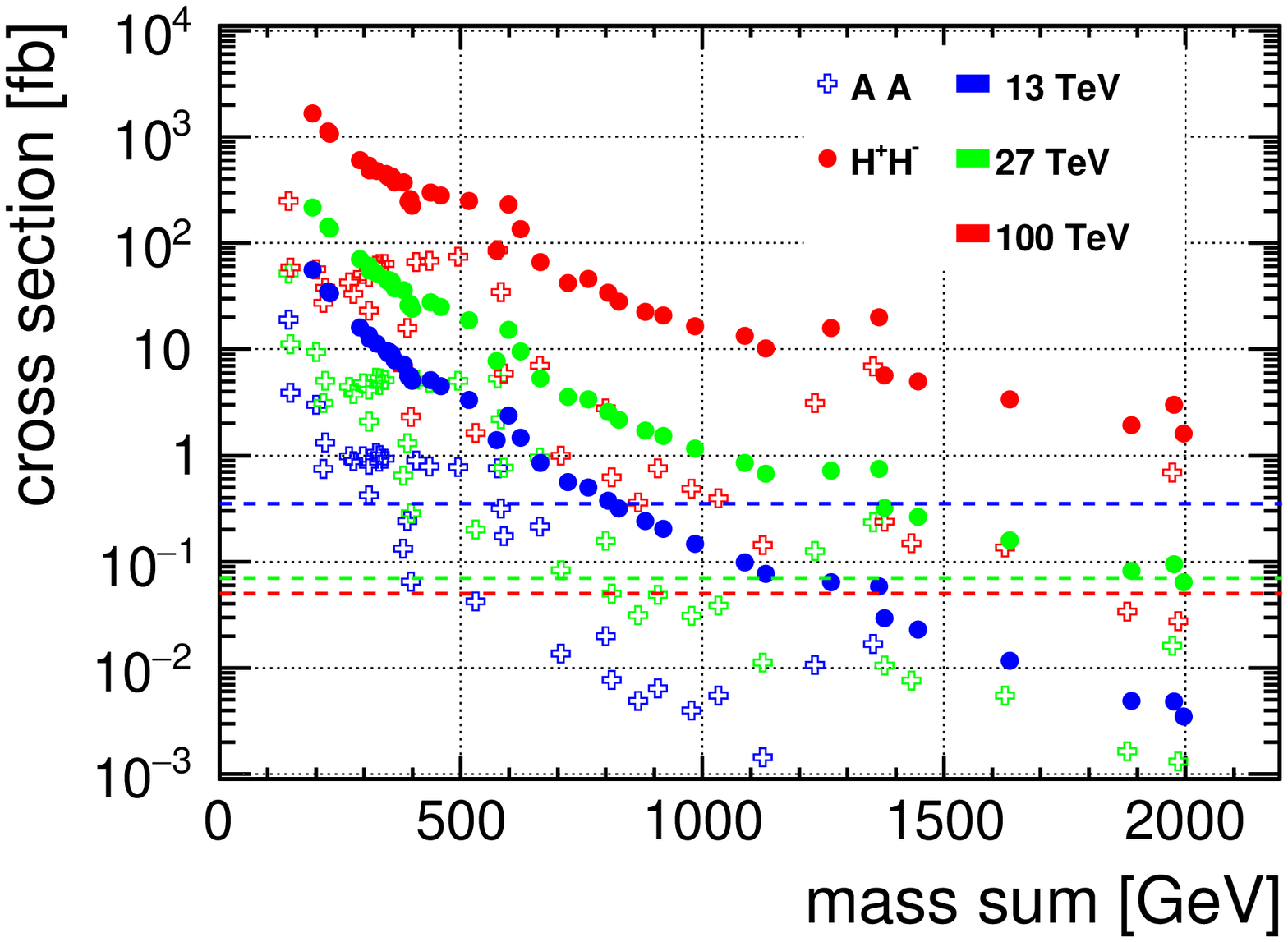}
\end{center}  
\caption{ Production cross sections for benchmarks from tables \ref{tab:bench} and \ref{tab:bmhigh} as a function of the produced scalar mass sum, for VBF production channels, for 13\,TeV, 27\,TeV and 100\,TeV proton collider options.
  Horizontal dashed lines indicate minimal cross sections required to produce 1000 events at the respective energy, assuming design luminosity.
\label{fig:crosvbf}}
\end{figure}
\begin{table}[p]
\small
\begin{center}
\begin{tabular}{|l|l|l|l||c|c|c||c|c|c|}
\hline 
No. & $M_H$ & $M_A$ & $M_{H^\pm}$ & $(AA)_{13} $  &$(AA)_{27}$&$(AA)_{100}$ &$ (H^+\,H^-)_{13}$ & $ (H^+\,H^-)_{27}$&$(H^+\,H^-)_{100}$\\ \hline \hline
\bf BP1 & 72.77 & 107.803 & 114.639 &0.750&3.11&27.5&33.5&137&1070\\ \hline
BP2 & 65 & 71.525 & 112.85 &18.9&52.0&250&34.8&142&1120\\ \hline
BP3 & 67.07 & 73.222 & 96.73 &3.87&11.1&58.9&55.5&217&1660 \\ \hline
BP4 & 73.68 & 100.112 & 145.728 &2.99&9.44&56.5&16.0&70.9&606 \\ \hline
\bf BP6 & 72.14 & 109.548 & 154.761 &1.33&5.02 &38.0&13.5 &61.4 &536 \\ \hline
BP7 & 76.55 & 134.563 & 174.367 &0.988 &4.43&42.5&9.22&43.5 &420 \\ \hline
\bf BP8 & 70.91 & 148.664 & 175.89  &0.982&4.78 &51.3&9.36&42.8 &419 \\ \hline
BP9 & 56.78 & 166.22 & 178.24  &1.01&5.34 &63.8&9.11&43.8 &424 \\ \hline
BP10 & 76.69 & 154.579 & 163.045  &0.819&4.17 &47.9&11.3&51.6 &478 \\ \hline
BP11 & 98.88 & 155.037 & 155.438   &0.422&2.08 &23.0&12.5& 55.3 &484 \\ \hline
BP12 & 58.31 & 171.148 & 172.96  &0.937&5.13 &63.7&9.66&45.5 &448 \\ \hline
BP13 & 99.65 & 138.484 & 181.321 &0.888&3.78&33.4&7.87&37.3&373 \\ \hline
\bf BP14 & 71.03 & 165.604 & 175.971 &0.875&4.63 &55.0 &9.36&44.9 &418 \\ \hline
\bf BP15 & 71.03 & 217.656 & 218.738 &0.788&4.84 &67.7 &5.12&27.7&298 \\ \hline
\bf BP16 & 71.33 & 203.796 & 229.092 &0.895&5.15 &66.0&4.49&25.0&280 \\ \hline
BP18 & 147 & 194.647 & 197.403 &0.240&1.30 &15.9&5.57&26.8 &258 \\ \hline
BP19 & 165.8 & 190.082 & 195.999 &0.133&0.646  &6.67&5.50&26.0&246 \\ \hline
BP20 & 191.8 & 198.376 & 199.721 &0.0653&0.284 &2.32&5.05&23.9& 223\\ \hline
\bf BP21 & 57.475 & 288.031 & 299.536 &0.756&5.29 &85.5&2.37 &15.2& 231\\ \hline
\bf BP22 & 71.42 & 247.224 & 258.382 &0.777 &5.04 &74.3&3.31&18.8&248 \\ \hline
BP23 & 62.69 & 162.397 & 190.822  &1.06&5.42&61.5 &7.21&35.7& 372\\ \hline
\end{tabular}

\caption{ \label{tab:VBF_low} Production cross sections for BPs from table \ref{tab:bench} in \fb~ for $X+\text{dijet}$ at proton-proton colliders for varying center-of-mass energies.  No VBF cuts were applied.}
\end{center}
\end{table}
\begin{table}[p]
\small
\begin{center}
\begin{tabular}{|l|l|l|l||c|c|c||c|c|c|}
\hline 
No. & $M_H$ & $M_A$ & $M_{H^\pm}$ & $(AA)_{13} $  &$(AA)_{27}$&$(AA)_{100}$ &$ (H^+\,H^-)_{13}$ & $ (H^+\,H^-)_{27}$&$(H^+\,H^-)_{100}$\\ \hline \hline
HP1 & 176 & 291.36 & 311.96 &0.319&2.20&34.4&1.47&9.57& 136\\ \hline
HP2 & 557 & 562.316 & 565.417 &0.00145&0.0112&0.143&0.0769&0.674&10.2 \\ \hline
HP3 & 560 & 616.32 & 633.48 &0.0107&0.126&3.12&0.0643&0.718&15.8 \\ \hline
HP4 & 571 & 676.534 & 682.54 &0.0169&0.234&6.90&0.0587&0.751& 19.9\\ \hline
HP5 & 671 & 688.108 & 688.437 &-&0.0105&0.239&0.0295&0.320&5.71 \\ \hline
HP6 & 713 & 716.444 & 723.045 &-&0.00762&0.149&0.0230&0.266&4.99 \\ \hline
HP7 & 807 & 813.369 & 818.001 &-&0.00553&0.137&0.0117&0.160&3.37 \\ \hline
HP8 & 933 & 939.968 & 943.787 &-&0.00165&0.0340&0.00492&0.0821&1.94 \\ \hline
HP9 & 935 & 986.22 & 987.975 &-&0.0162&0.696&0.00482&0.0945&3.00 \\ \hline
\bf HP10 & 990 & 992.36 & 998.12 &-&0.00131&0.0275&0.00349&0.0640&1.61 \\ \hline
HP11 & 250.5 & 265.49 & 287.226 &0.0423&0.200&1.64&1.39&7.76&84.4 \\ \hline
HP12 & 286.05 & 294.617 & 332.457 &0.175&0.772&5.89&0.853&5.29& 66.0\\ \hline
HP13 & 336 & 353.264 & 360.568 &0.0136&0.0829&1.00&0.565&3.56&42.1 \\ \hline
HP14 & 326.55 & 331.938 & 381.773 &0.216&0.959&7.00&0.499&3.38&45.7  \\ \hline
HP15 & 357.6 & 399.998 & 402.568 &0.0200&0.157&2.77&0.375&2.57&34.0 \\ \hline
HP16 & 387.75 & 406.118 & 413.464 &0.00768&0.0498&0.622&0.318&2.18&28.0 \\ \hline
HP17 & 430.95 & 433.226 & 440.624 &0.00490&0.0316&0.361&0.240&1.72&22.5 \\ \hline
HP18 & 428.25 & 453.979 & 459.696 &0.00640&0.0486&0.756&0.203&1.52&20.8 \\ \hline
HP19 & 467.85 & 488.604 & 492.329 &0.00396&0.0311&0.484&0.148&1.17&16.4 \\ \hline
HP20 & 505.2 & 516.58 & 543.794 &0.00550&0.0388&0.397&0.0982&0.852&13.3 \\ \hline
\end{tabular}

\caption{ \label{tab:VBF_high} Production cross sections for HPs from table \ref{tab:bmhigh} in \fb~ for $X+\text{dijet}$ at proton-proton colliders for varying center-of-mass energies. No VBF cuts were applied.}
\end{center}
\end{table}
Note that we did not apply VBF-like cuts, as, depending on the parameter point, different channels contribute; for $AA$ production, this can e.g. be gluon-gluon or vector-boson fusion to $h$ with successive decays to $AA$, as well as diagrams with e.g. a charged scalar in the $t-$channel. For $H^+\,H^-$ final states, standard dijet production with $Z/ \gamma$ radiation with successive decay into $H^+\,H^-$ can also play a significant role.

For $AA$ production, comparing to non-VBF like topologies, we encounter enhancement rates up to three orders of magnitude when considering the VBF-like contribution, especially e.g. for HP14 and HP20, where the largest relative growth takes place for 13 \TeV. However, at this center-of-mass energy the total rate remains small. If we consider accessible points only, with at least 1000 events being produced over the full run in the VBF mode, the largest enhancement can be seen for HP20 at 100 \TeV and HP14 at 27 \TeV or 100 \TeV, where the production cross section increases by roughly three orders of magnitude. A detailed analysis for the latter point shows that the predominant contribution for this point at e.g. 100 \TeV~ stems from off-shell $H^+ A$ production and subsequent decay $H^+\,\rightarrow\,W^+ A$ as well as processes decribed by diagrams with a charged scalar in the $t$-channel, initiated by $WW$ fusion.\footnote{\label{foot:gauge} Note this identification stems from graph identification within Madgraph5; in general, only the complete set of $WW$ initiated diagrams is gauge-invariant. The above statement has been derived by evaluations in the unitary gauge.} Enhancements by more than an order of magnitude are also observed for HPs 11 and 12 at the same center-of-mass energies 27\,TeV and 100\,TeV, HPs 2 and 16 at 100 \TeV, followed by BP\,2 accessible already in Run II and BP 20 at 100 TeV. At 13 \TeV, the cross section for BP\,2 rises from $\sim\,1.5\, \fb~$ to $\sim\,19\,\fb$ when VBF-like topologies are considered. This can again be traced back mainly to contributions from $H^\pm\,A$ production with successive decays $H^\pm\,\rightarrow\,A\,W^\pm$. At 13 \TeV, for example, BPs 11 and 13 might now be accessible at the HL-LHC in the AA VBF channel. At 27 \TeV, BP20 as well as 7 additional HPs might now be visible; at 100 \TeV, nearly all HPs have large enough cross sections in this channel, with only HPs 8 and 10 having cross sections $\lesssim\,0.04\, \fb$. We show the enhancement for points with more than 1000 events with full integrated luminosity in Fig.~\ref{fig:ratioaa}.\\
\begin{figure}[tb]
\begin{center}
  \includegraphics[width=0.6\textwidth]{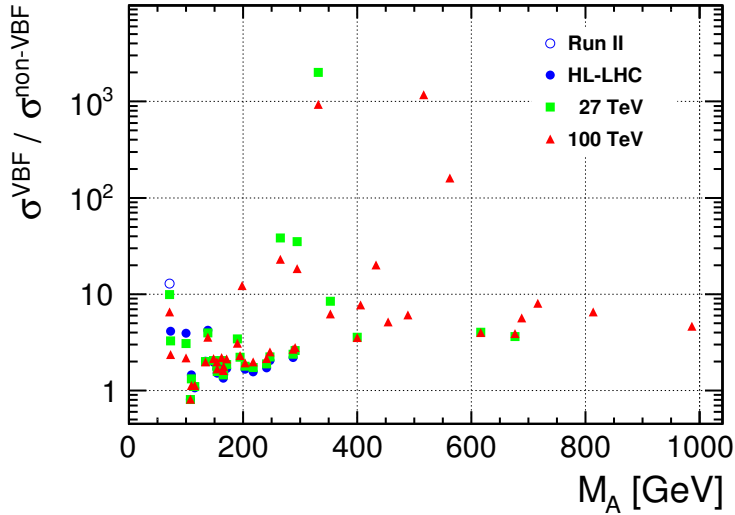}
\end{center}  
\caption{ Enhancement of $AA$ production cross sections at pp colliders with various center-of-mass energies when VBF-type topologies are included. Only points with minimal cross-section requirements as specified in the text are shown. See detailed discussion in main body of paper. 
\label{fig:ratioaa}}
\end{figure}
In Fig.~\ref{fig:crosvbf}, although in general a decrease in the cross sections is observed for rising masses, there are points which deviate from this behaviour, as e.g. the production cross sections for $AAjj$ at 100\,\TeV. As an example, HP4 here leads to a cross section of about 7 \fb~, while the production cross section for HP5 is more an order of magnitude lower, while masses of $A$ are quite similar. This can be traced back to the production of an off-shell $h$ with two jets, where the $h$ subsequently decays to $AA$. This process is mediated via the $\bar{\lam}_{345}$ coupling, which grows with the difference between $M_H^2$ and $M_A^2$. In fact, concentrating on the dominant contribution, namely $g g\,\rightarrow\,h^{*}\,g\,g$, with subsequent decays $h^{*}\,\rightarrow\,A\,A$, we find that $\lb {\frac{\bar{\lam}_{345, HP5}}{\bar{\lam}_{345, HP4}}}\rb^2\,\sim\,0.026$. The production cross sections in this mode are $4.7\fb$ and $0.11\fb$ {for HP4 and HP5}, respectively, displaying the same ratio. Additional contributions in both points stem from VBF diagrams with e.g. a charged or neutral scalar in the $t$-channel; for HP 4/ 5, these contribute roughly $4\%\,/10\%$ to the total cross section.

For the $H^+\,H^-$ channel, the VBF-induced cross sections are up to a factor of 2 larger than for the direct production;  maximal enhancement is observed for HP4 at 100 \TeV. In fact, enhancements can mainly present for this collider option. In contrast, e.g. for BP3 at 13 \TeV, the VBF-type cross section only amounts to about $20\%$ of the direct production.
As before, we note a general decrease of the cross sections as masses rise. However, we can again observe that for similar mass scales, there can be exceptions where cross sections differ by about a factor 3. Again, this can be traced back to diagrams that are mediated via the SM-like scalar $h$. The coupling between $h$ and $H^+H^-$ is given by
\begin{\eqn*}
\lam_3\,\equiv\,\lam_{345}-2\frac{M_{H}^2-M_{H^\pm}^2}{v^2}.
\end{\eqn*}
As an example, we can consider production cross sections for BP21 and HP11 at 100 \TeV; both points feature similar charged scalar masses, however, the mass differences to the dark matter candidate vary largely. For BP21, we have $\lam_3\,\sim\,2.9$, while for HP11, the corresponding value is given by $\lam_3\,\sim\,0.5$. This leads to a relative factor of around 30 for contributions which are triggered by $h$-exchange in the $s$-channel. In fact, the corresponding cross sections stemming from gluon-fusion are 118 \fb~ for BP21 and 4 \fb~ for HP11, reflecting this ratio. Other diagrams come from $p\,p\,\rightarrow\,j\,j\,\gamma\,(Z)$, with the electroweak gauge boson decaying into the charged scalars, as well as diagrams with charged scalars in the $t$-channel. Due to quantum interference, it is not obvious to disentangle these from $h$-induced contributions. However, for HP11 it can be stated that gg-induced processes contribute roughly $\sim\,5\%$ to the total cross section, while the corresponding number for BP21 is $\gtrsim\,50\%$. Similarly, one can compare cross sections for HP4 and HP5 a 100 \TeV~ center-of-mass energy; although these points feature similar charged scalar masses, the cross sections differ by a factor 3.5. This can again be traced back to differences in $\lam_3$, which is given by 4.15/ 0.64 for HP4 / HP5, respectively. Comparing numbers from gg-induced processes only, which are dominantly mediated via $h$-exchange, we find that the cross sections are given by 10.5 \fb~ and 0.229 \fb~ respectively, representing the above hierarchy in the coupling. In other channels, processes which are $h$-mediated are contributing mainly for HP4. As before, a clear disentanglement is not possible due to interference effects, however, one can state that for HP4 at least $50 \%$ of the total cross section are mediated via $h$, while this number goes down to about $4 \%$ for HP5. 

In summary, for $AA$ final states inclusion of processes with additional jets can greatly improve the collider reach. For $H^+\,H^-$, instead, maximal enhancements reach a factor 2 at a 100 \TeV~ collider, while for lower center-of-mass energies the respective cross sections can be up to a factor 5 smaller than the direct production channel.

\subsection{Purely photon-induced processes}

We also briefly comment on the possibility of {observing} photon-induced {production} processes {using} forward proton spectrometers, as e.g. discussed in \cite{Adamczyk:2017378,Albrow:1753795,CMS:2021ncv}. Here, photons are emitted from the protons, and the final state $p'\,p'+\,X$ is measured, with $X$ being the final state generated via photon-fusion {and $p'$ denoting intact protons in the final state, which could be measured in the proton spectrometers.} For the IDM, the only possible process into novel final states is given by
\begin{\eqn*}
p\,p\,\rightarrow\,p'\,p'\,H^+\,H^-,
\end{\eqn*}
as no other BSM final state can be generated via photon-photon fusion at tree-level\footnote{{In principle processes would be possible via the photon-photon-Higgs vertex, possibly allowing for $AA$ photo-production, and also contribute to the above process, albeit at higher order. This is currently not implemented in our framework and beyond the scope of the current work.} }. We present the production cross-sections for all benchmark scenarios in tables \ref{tab:aabp} and \ref{tab:aahp}, respectively.
\begin{table}[p]
\small
\begin{center}
\begin{tabular}{|l|l|l|l|c|c|}
\hline 
No. & $M_H$ & $M_A$ & $M_{H^\pm}$ & $ (H^+\,H^-)_{13}$ & $ (H^+\,H^-)_{100}$\\ \hline
\bf BP1 & 72.77 & 107.803 & 114.639 &0.404&2.63\\ \hline
BP2 & 65 & 71.525 & 112.85 &0.425&2.74\\ \hline
BP3 & 67.07 & 73.222 & 96.73 &0.695&4.14\\ \hline
BP4 & 73.68 & 100.112 & 145.728 &0.184&1.37\\ \hline
\bf BP6 & 72.14 & 109.548 & 154.761 &0.150&1.16\\ \hline
BP7 & 76.55 & 134.563 & 174.367 &0.100&0.833 \\ \hline
\bf BP8 & 70.91 & 148.664 & 175.89  &0.0971&0.817 \\ \hline
BP9 & 56.78 & 166.22 & 178.24  & 0.0927&0.786 \\ \hline
BP10 & 76.69 & 154.579 & 163.045  &0.126&1.00\\ \hline
BP11 & 98.88 & 155.037 & 155.438& 0.148  &1.15\\ \hline
BP12 & 58.31 & 171.148 & 172.96  &0.103&0.855 \\ \hline
BP13 & 99.65 & 138.484 & 181.321 &0.0875&0.750 \\ \hline
\bf BP14 & 71.03 & 165.604 & 175.971 &0.0970&0.814 \\ \hline
\bf BP15 & 71.03 & 217.656 & 218.738 &0.0455&0.440 \\ \hline
\bf BP16 & 71.33 & 203.796 & 229.092 &0.0384&0.388 \\ \hline
BP18 & 147 & 194.647 & 197.403 &0.0651&0.592 \\ \hline
BP19 & 165.8 & 190.082 & 195.999 &0.0667&0.604 \\ \hline
BP20 & 191.8 & 198.376 & 199.721 &0.0625&0.572 \\ \hline
\bf BP21 & 57.475 & 288.031 & 299.536 &0.0143&0.181 \\ \hline
\bf BP22 & 71.42 & 247.224 & 258.382 &0.0248&0.277 \\ \hline
BP23 & 62.69 & 162.397 & 190.822  &0.0734&0.651 \\ \hline
\end{tabular}

\caption{ \label{tab:aabp} Production cross sections for BPs from table \ref{tab:bench} in \fb~ for $X p' p'$ at a 13 and 100 \TeV pp collider.  {No cuts are applied on the scattered proton kinematics.}}
\end{center}
\end{table}
\begin{table}[p]
\small
\begin{center}
\begin{tabular}{|l|l|l|l|c|c|}
\hline 
No. & $M_H$ & $M_A$ & $M_{H^\pm}$ & $ (H^+\,H^-)_{13}$ & $ (H^+\,H^-)_{100}$\\ \hline
HP1 & 176 & 291.36 & 311.96 &0.0122&0.161\\ \hline
HP2 & 557 & 562.316 & 565.417 &0.00106&0.0276 \\ \hline
HP3 & 560 & 616.32 & 633.48 &-&0.0194 \\ \hline
HP4 & 571 & 676.534 & 682.54 &-&0.0154 \\ \hline
HP5 & 671 & 688.108 & 688.437 &-&0.0150 \\ \hline
HP6 & 713 & 716.444 & 723.045 &-&0.0128\\ \hline
HP7 & 807 & 813.369 & 818.001 &-&0.00868\\ \hline
HP8 & 933 & 939.968 & 943.787 &-&0.00547\\ \hline
HP9 & 935 & 986.22 & 987.975 &-&0.00471 \\ \hline 
\bf HP10 & 990 & 992.36 & 998.12 &-&0.00457 \\ \hline
HP11 & 250.5 & 265.49 & 287.226 &0.0167&0.205 \\ \hline
HP12 & 286.05 & 294.617 & 332.457 &0.00958&0.134\\ \hline
HP13 & 336 & 353.264 & 360.568 &0.00699&0.106 \\ \hline
HP14 & 326.55 & 331.938 & 381.773 &0.00557&0.0895\\ \hline
HP15 & 357.6 & 399.998 & 402.568 &0.00450&0.0766\\ \hline
HP16 & 387.75 & 406.118 & 413.464 &0.00403&0.0709 \\ \hline
HP17 & 430.95 & 433.226 & 440.624 &0.00310&0.0587 \\ \hline
HP18 & 428.25 & 453.979 & 459.696 &0.00261&0.0514 \\ \hline
HP19 & 467.85 & 488.604 & 492.329 &0.00195&0.0420 \\ \hline
HP20 & 505.2 & 516.58 & 543.794 &0.00127&0.0310 \\ \hline
\end{tabular}

\caption{ \label{tab:aahp} Production cross sections for HPs from table \ref{tab:bmhigh} in \fb~ for $X p' p'$ at a 13 and 100 \TeV pp collider.  {No cuts are applied on the scattered proton kinematics.}}
\end{center}
\end{table}
{No cuts on the scattered proton kinematics are applied.}
As for direct pair-production, the cross sections are determined by the available phase space, given by the masses of the charged scalars, and exhibit decline with rising mass scales.
The production cross sections are lower by factors 300 for 13 \TeV and up to 800 for 100 \TeV with respect to the direct pair-production cross sections, given in tables \ref{tab:xs13low}, \ref{tab:xs13high} and \ref{tab:xs100low}, \ref{tab:xs100high} for 13 \TeV and 100 \TeV, respectively. Therefore, all points here would in principle be within reach first in direct pair-production, using again our simple counting criterium. The {photon-fusion} mode would in principle provide an additional test of the model for photon induced processes. {However, even not taking into account the acceptance of the proton spectrometers} only BPs 1-3 would be accessible {at the HL-LHC}, corresponding to a mass range up to 200 \GeV for the sum of the produced particles. At 100 \TeV, all low mass points as well as HPs 1, 11-18 would be accessible, enhancing the mass range to 900 \GeV.

\subsection{Muon collider}

Recently, discussions of a muon collider have again raised some interest in the community (see e.g. \cite{mu_2707}). We therefore present cross sections at such a collider for two collider options, namely, for center-of-mass energies of $10\,\TeV$ and $30\,\TeV$\footnote{Note that without taking beamstrahlung and initial state radiation effects into account, the cross sections for $\mu^+\mu^-$ and $e^+e^-$ induced processes are the same as lepton masses are negligible for the considered center-of-mass energies. Since beamstrahlung and initial state radiation effects are much less important for $\mu^+\mu^-$ they were therefore not taken into account in the presented study.}.

For direct production, we found that cross sections are similar for all BPs and HPs, given by $0.13\,\fb$ for $HA$ production and $0.31\,\fb$ for $H^+H^-$ production at the 10 \TeV~ collider, respectively; cross sections at 30 \TeV~ are about an order of magnitude lower\footnote{Cross sections might slightly rise due to radiative return (see e.g. \cite{Chakrabarty:2014pja}).}. We therefore list results for VBF-type production modes only; in particular, we consider
\begin{\eqn*}
  \mu^+\,\mu^-\,\rightarrow\,\nu_\mu\,\bar{\nu}_\mu A A,\;\;\;
  \mu^+\,\mu^-\,\rightarrow\,\nu_\mu\,\bar{\nu}_\mu H^+ H^-.
\end{\eqn*}
Production cross sections for these processes can be found in tables \ref{tab:mumulow} and  \ref{tab:mumuhigh}, and are compared in Fig.~\ref{fig:crosmumu}.
\begin{figure}[tb]
\begin{center}
  \includegraphics[width=0.6\textwidth]{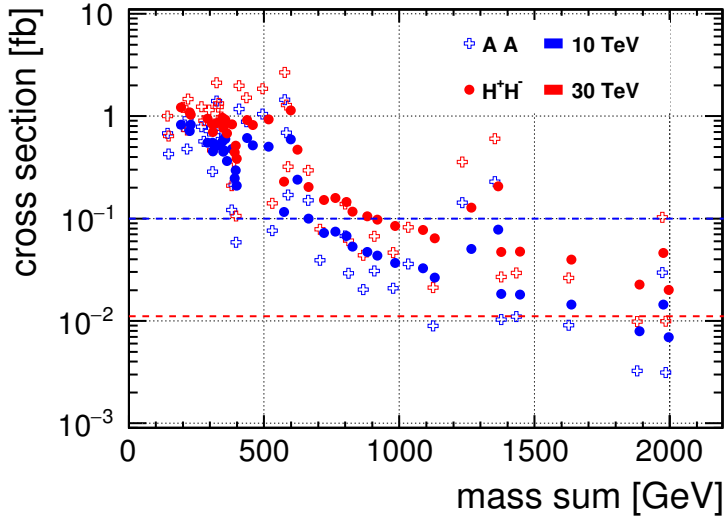}
\end{center}  
\caption{ Production cross sections for benchmarks from tables \ref{tab:bench} and \ref{tab:bmhigh} as a function of the produced scalar mass sum, for $AA$ and $H^+H^-$ production at 10\,TeV and 30\,TeV muon collider.
Horizontal dashed lines indicate minimal cross sections required to produce 1000 events at the respective energy, assuming 5 year integrated design luminosity.
\label{fig:crosmumu}}
\end{figure}
Depending on the parameter point, different diagrams contribute. For the low-mass BPs, production cross sections range between 0.06 \fb~ and 1.17 \fb~ at 10 \TeV~ and between 0.1 \fb~ and 3 \fb~ at 30 \TeV. For example, the dominant contribution to the cross section for $AA$ final states at BP21, the benchmark point with the highest rates, stems from diagrams with a charged scalar in the $t$-channel (see footnote \ref{foot:gauge}). For high-mass HPs, cross sections start basically an order of magnitude lower, and can reach up to roughly 1 \fb~ at both center-of-mass energies, depending on the benchmark point and production mode. Note that for $H^+H^-$ the VBF-like production almost always renders rates higher than direct pair-production, with the exception of the HPs at 10 \TeV~ center-of-mass energy. For example, for BP3 at 30 \TeV, diagrams with W-boson fusion to a Z-boson or photon with successive decay to $H^+\,H^-$ are predominant, with slightly lower contributions from diagrams with an $A$ or $H$ in the t-channel.

\begin{table}[p]
\small
\begin{center}
\begin{tabular}{|l|l|l|l|c|c|c|c|}
\hline 
No. & $M_H$ & $M_A$ & $M_{H^\pm}$ & $(AA)_{10}$  & $ (H^+\,H^-)_{10}$ &  $(AA)_{30}$  & $ (H^+\,H^-)_{30}$\\ \hline
\bf BP1 & 72.77 & 107.803 & 114.639 &0.476&0.828&0.732&1.03\\ \hline
BP2 & 65 & 71.525 & 112.85 & 0.678&0.714&1.00&1.09\\ \hline
BP3 & 67.07 & 73.222 & 96.73 &0.426&0.825&0.641&1.22\\ \hline
BP4 & 73.68 & 100.112 & 145.728 &0.798&0.549&1.22&0.946 \\ \hline
\bf BP6 & 72.14 & 109.548 & 154.761 &0.954&0.551&1.47&0.834 \\ \hline
BP7 & 76.55 & 134.563 & 174.367 &0.787 &0.451&1.24&0.767 \\ \hline
\bf BP8 & 70.91 & 148.664 & 175.89  &0.714&0.463&1.15&0.807 \\ \hline
BP9 & 56.78 & 166.22 & 178.24  &0.813 &0.600&1.33&0.921 \\ \hline
BP10 & 76.69 & 154.579 & 163.045  &0.521&0.515&0.837&0.857 \\ \hline
BP11 & 98.88 & 155.037 & 155.438   &0.286&0.452&0.464&0.698 \\ \hline
BP12 & 58.31 & 171.148 & 172.96  &0.600&0.564&0.984&0.969 \\ \hline
BP13 & 99.65 & 138.484 & 181.321 &0.567 &0.366&0.904&0.680 \\ \hline
\bf BP14 & 71.03 & 165.604 & 175.971 &0.754 &0.601&1.23&0.860 \\ \hline
\bf BP15 & 71.03 & 217.656 & 218.738 &0.880 &0.610&1.51&0.919 \\ \hline
\bf BP16 & 71.33 & 203.796 & 229.092 &1.17 &0.519&1.98&0.817 \\ \hline
BP18 & 147 & 194.647 & 197.403 &0.214 &0.296&0.362& 0.515\\ \hline
BP19 & 165.8 & 190.082 & 195.999 &0.121 &0.247&0.209&0.442 \\ \hline
BP20 & 191.8 & 198.376 & 199.721 &0.0586 &0.210&0.106&0.385 \\ \hline
\bf BP21 & 57.475 & 288.031 & 299.536 &1.45 &0.595&2.67&1.14 \\ \hline
\bf BP22 & 71.42 & 247.224 & 258.382 &1.05 &0.504&1.85&0.931 \\ \hline
BP23 & 62.69 & 162.397 & 190.822  &1.41 &0.478&2.13&0.832 \\ \hline
\end{tabular}

\caption{ \label{tab:mumulow} Production cross sections for BPs from table \ref{tab:bench} in \fb~ for $X \nu_\mu \bar{\nu}_\mu$ at a 10 and 30 \TeV~ muon-collider. }
\end{center}
\end{table}
\begin{table}[p]
\small
\begin{center}
\begin{tabular}{|l|l|l|l|c|c|c|c|}
\hline 
No. & $M_H$ & $M_A$ & $M_{H^\pm}$ & $(AA)_{10}$  & $ (H^+\,H^-)_{10}$ &  $(AA)_{30}$  & $ (H^+\,H^-)_{30}$\\ \hline
HP1 & 176 & 291.36 & 311.96 &0.691&0.240&1.28&0.470 \\ \hline
HP2 & 557 & 562.316 & 565.417 &0.00892&0.0265&0.0211&0.0639 \\ \hline
HP3 & 560 & 616.32 & 633.48 &0.143 &0.0506&0.356& 0.128\\ \hline
HP4 & 571 & 676.534 & 682.54 &0.229 &0.0777&0.602&0.206 \\ \hline
HP5 & 671 & 688.108 & 688.437 &0.0103&0.0184&0.0268&0.0473 \\ \hline
HP6 & 713 & 716.444 & 723.045 &0.0110 &0.0181&0.0295&0.0474 \\ \hline
HP7 & 807 & 813.369 & 818.001 &0.00910 &0.0144&0.0263&0.0397 \\ \hline
HP8 & 933 & 939.968 & 943.787 &0.00324&0.00792&0.00986&0.0226 \\ \hline
HP9 & 935 & 986.22 & 987.975 &0.0297 &0.0144&0.103&0.0460 \\ \hline
\bf HP10 & 990 & 992.36 & 998.12 &0.00312 &0.00695&0.00991&0.0201 \\ \hline
HP11 & 250.5 & 265.49 & 287.226 &0.0760 &0.116&0.141&0.230 \\ \hline
HP12 & 286.05 & 294.617 & 332.457 &0.170 &0.0996&0.320&0.204 \\ \hline
HP13 & 336 & 353.264 & 360.568 &0.0392&0.0722&0.0785&0.152 \\ \hline
HP14 & 326.55 & 331.938 & 381.773 &0.151&0.0746&0.295&0.159 \\ \hline
HP15 & 357.6 & 399.998 & 402.568 &0.0677 &0.0678&0.139&0.145 \\ \hline
HP16 & 387.75 & 406.118 & 413.464 &0.0291 &0.0533&0.0609&0.117 \\ \hline
HP17 & 430.95 & 433.226 & 440.624 &0.0203&0.0473&0.0438&0.105 \\ \hline
HP18 & 428.25 & 453.979 & 459.696 &0.0308 &0.0435&0.0669&0.0979 \\ \hline
HP19 & 467.85 & 488.604 & 492.329 &0.0208 &0.0368&0.0464&0.0849 \\ \hline
HP20 & 505.2 & 516.58 & 543.794 &0.0359 &0.0326&0.0824&0.0773 \\ \hline
\end{tabular}

\caption{ \label{tab:mumuhigh} Production cross sections for HPs from table \ref{tab:bmhigh} in \fb~ for $X \nu_\mu \bar{\nu}_\mu$ at a 10 and 30 \TeV~ muon-collider.}
\end{center}
\end{table}

As before, in general one can observe a decrease of production cross sections with rising mass scales, where however some exceptions exist. For $H^+H^-$ production, it is again instructive to compare cross sections for BP21 and HP11, which feature similar charged scalar masses but different $M_H$, leading to a factor 5 difference in production cross sections at 30 \TeV. This difference can be traced back to the interference between two different gauge-invariant sets of diagrams which contribute to this process, with $W^+W^-$ and $W\mu$ fusion, which we label GI I and GI II, respectively; the two sets of diagrams are displayed in Appendix \ref{app:diags}. Contributions from these sets of diagrams are shown in table \ref{tab:comp} where we also consider two additional parameter points HP11b, HP11c which have the same charged or charged and dark matter mass as BP21. We see that, while contributions to GI II mainly depend on the masses of the charged scalars, in GI~I the masses of the neutral dark scalars also play a role via diagrams with these particles in the $t$-channel. 
\begin{center}
\begin{table}
\begin{center}
\begin{tabular}{|c|c|c|c|c|}
\hline
&BP21&HP11&HP11b&HP11c\\ \hline
$M_{H^+}$&299.536&287.226&299.536&299.536\\
$M_{H}$&57.475&250.5&250.5&57.475\\
$M_A$&288.031&265.49&265.49&265.49\\ \hline
GI I&19.01(7)&19.69(8)&18.04(6)&18.76(7) \\
GI II&17.89(5)&19.43(5)&17.83(5)&17.86(6) \\ \hline
Eqn (\ref{eq:vbfmumuhphm})&1.12 (9)&0.23(9)&0.21(7)&0.90(9) \\
GI I + GI II&1.129(3)&0.2293(6)&0.2274(7)& 0.973(3)\\ \hline
total&1.144(5)&0.2297(7)&0.2276(8)&0.968(3) \\ \hline
\end{tabular}
\caption{\label{tab:comp} Transition between BP21 and HP11, including different contributions from gauge-invariant sets of diagrams with $WW$ (GI I) and $W\mu$ (GI II) fusion, for $H^+H^-$ production in the VBF-like mode at a muon collider with 30 \TeV~ center-of-mass energy. Masses are in \GeV~ and cross sections in \fb. See text for details.}
\end{center}
\end{table}
\end{center}

From the table, we observe that the final contribution seems to dominantly stem from a fine-tuned cancellation between these two type of diagrams according to
\begin{\eqn}\label{eq:vbfmumuhphm}
\int_{PS} |\mathcal{M}_I+\mathcal{M}_{II}|^2\,\simeq\,\int_{PS}\,\lb |\mathcal{M}_I|^2-|\mathcal{M}_{II}|^2\rb
\end{\eqn}
where $\mathcal{M}_i\,=\,|\mathcal{M}_{i}|\,e^{i\,\varphi_i}$ and $\int_{PS}$ denotes integration over phase space. The above equation is e.g. fulfilled if the integrated matrix elements differ by a phase and obey
\begin{\eqn*}
\int_{PS} |\mathcal{M}_{II}|^2\,\simeq\,-\int_{PS} |\mathcal{M}_{I}||\mathcal{M}_{II}|\cos\lb\varphi_{II}-\varphi_{I} \rb,
\end{\eqn*}
A similar observation can be done comparing HP4 and HP5, which feature similar scalar masses, but vary in the mass differences between $M_{H^+}$ and $M_H$. A detailed study shows that, as before, a larger mass gap increases contributions from GI I, therefore leading to a larger total result.\\

For the $AA$ channel, things are slightly different. Here, the main contribution stems from $WW$ fusion only, where the corresponding diagrams can be found in appendix \ref{app:diagsaa}. It is instructive to consider the contributions triggered by $h$- exchange, with the coupling strength $\bar{\lam}_{345}$ (cf. eq. (\ref{eq:l345b}) ), with respect to the remaining diagrams\footnote{Note that this discussion now assumes unitary gauge; in general, the above split is not gauge-invariant.}. We again consider HP4 and HP5 at 30 \TeV; these points have similar charged and heavier neutral scalar masses, but largely different $M_H$, resulting in different $\bar{\lam}_{345}$ values. We list the separate contributions in table \ref{tab:hp4hp5aa}. Note that the total contribution is dominated by the $h-$exchange diagram for HP4, corresponding to the relatively large $\bar{\lam}_{345}$ value. From the table, we can see that indeed the different terms are found to be proportional to the ratio of the $hH^+H^-$ coupling squared. 
\begin{center}
\begin{table}
\begin{center}
\begin{tabular}{|c|c|c|} \hline
&HP4&HP5\\ \hline
$M_A$&676.534&688.108\\
$M_{H^+}$&682.54&688.437\\
$M_{H}$&571&671\\
$\bar{\lam}_{345}$&3.88&0.62\\ \hline
h-exchange only&0.510(2)&0.01259(4)\\
all others&0.01625(4)&0.01347(5)\\
interference&0.078(4)&0.0008(1)\\ \hline
total&0.604(3)&0.02686(9)\\ \hline
\end{tabular}
\caption{\label{tab:hp4hp5aa} Comparison of different contributions to $AA$ final state in the VBF-like production mode at a muon collider with the center-of-mass energy of 30 \TeV. Masses are given in \GeV and cross sections in \fb.}
\end{center}
\end{table}
\end{center}
In general, however, the total contribution depends on all three dark scalar masses. This can be seen in table \ref{tab:bp21hp11aa}, where we compare BP21, HP12, as well as variations around BP21 where we vary the mass differences between the dark scalars.
Comparing BP21b and BPP21c, we observe that the contribution from $h$-exchange is directly proportional to $\bar{\lam}^2_{345}$, as expected. For a similar mass range of $M_A$, we can therefore tune the total cross section by at least an order of magnitude by varying the other dark scalar masses.
\begin{center}
\begin{table}
\begin{center}
\begin{tabular}{|c|c|c|c|c|} \hline
&BP21&HP12&BP21b&BP21c\\ \hline
$M_A$&288.031&294.617&288.031&288.031\\
$M_{H^+}$&299.536&332.457&329.536&329.536\\
$M_{H}$&57.475&286.05&57.475&280.031\\
$\bar{\lam}_{345}$&2.63&0.277&2.63&0.152\\ \hline
h-exchange only&2.25(1)&0.2155(8)&2.067(8)&0.00689(3)\\
all others&0.00706(2)&0.1899(5)&0.2223(7)&0.2216(8)\\
interference&0.36(1)&0.1094(9)&1.19(1)&0.067(1)\\ \hline
total&2.68(1)&0.3208(8)&3.48(1)&0.295(1)\\ \hline
\end{tabular}
\caption{\label{tab:bp21hp11aa} Comparison of different contributions to $AA$ final state in the VBF-like production mode at a muon collider with the center-of-mass energy of 30 \TeV. Masses are given in \GeV~ and cross sections in \fb.}
\end{center}
\end{table}
\end{center}

In \cite{Delahaye:2019omf}, the authors give a rough estimate of integrated luminosity that could be achieved at a muon collider, as a function of the center-of-mass energy. In particular
\begin{\eqn*}
\int\mathcal{L}\,\sim\,\lb\frac{ \sqrt{s}}{10\,\TeV} \rb^2.
\end{\eqn*}
For a 10 \TeV~ collider, they estimate an integrated lumosity of $10\,\ab^{-1}$ for a 5-year run. Applying the above expression for the higher center-of-mass energy, we roughly expect the integrated luminosity to be larger by one order of magnitude.

The authors equally state that target processes at 10 \TeV~ should have cross sections of $\mathcal{O}\lb \fb\rb$, with a similar rescaling at 30 \TeV. Using this criterium, we see that at 10 \TeV only BPs 16,21,22, {and 23} would be accessible in the VBF-like production of $AA$, while none of the HPs can be tested. At 30 \TeV, all low mass BPs are accessible; in the high mass region, HPs 2,5-8, 10 and 18-20 render too low cross sections. This corresponds to a maximal mass range of about $\sum_i M_i\,=\,1400\,\GeV$.

In accordance with the previous discussion, we can again alternatively require that at least 1000 events are produced in order to assess accesibility of a certain benchmark point. Using this criterium, all low mass BPs would be accessible during a 5 year run at 10 \TeV in all channels, with the exception of $AA$ production for BP20; this channel however provides a large enough cross section at 30 \TeV. For the high-mass HPs, HPs 1,3,4, 12 and 14 would be accessible in the $AA$ channel, where HP 1 and 11 have large enough cross sections in the $H^+H^-$ channel. This corresponds to a mass range of up to 600 \GeV (1400 \GeV) in the $H^+H^-$ ($AA$) channel. At 30 \TeV, all HPs would be accessible.

We want to emphasize again that the accessibility criterium of 1000 generated events can only be regarded as a first approximation and was introduced for comparison only; obviously, detailed investigations are needed in order to determine the true discovery range. We however consider this an easy selection criterium. More detailed results for investigation and reachability of the discussed benchmarks scenarios at CLIC can be found e.g. in \cite{Kalinowski:2018kdn,deBlas:2018mhx,Zarnecki:2020swm}.

\section{Conclusions}

We have presented several benchmarks for the Inert Doublet Model, a Two Higgs Doublet Model with a dark matter candidate, and provided predictions for the pair-production of dark scalars at the 13 \TeV~LHC, a high-energy upgrade, as well as a possible 100 \TeV~ proton-proton collider. We also gave predictions for pair-production cross sections at possible $\mu^+\mu^-$ colliders with various center-of-mass energies. Applying a simple counting criterium, we categorize {these} benchmarks in terms of their possible accessibility at different facilities. For example, after the high-luminosity run of the LHC, assuming target luminosity, for the low BPs in table \ref{tab:bench} all channels should be accessible, apart from the $AA$ final state which is suppressed due to small absolute values of the coupling $\lam_{345}$. Taking additionally VBF-like topologies {into account} for this final state then renders all but BPs 18 - 20 accessible after the HL-LHC run.  For the high benchmark points HPs 1 and {11-17} should be accessible in all but the $AA$ channel; for HPs {18 and 19}, in addition the $A\,H^{-},\,H\,H^{-}$ production modes render lower cross sections. For HP20, only the $H H^+$ {and $A H^+$} channel{s} look feasible. This corresponds to a possible reach up to about 500 \GeV~ for scalar masses. For $AA$, masses up to $300\,\GeV$ render large enough cross sections.

In turn, several possible future scenarios are considered: a high-energy upgrade to a center of mass energy of 27 \TeV, a 100 \TeV proton-proton facility, as well as a possible muon collider with different energy stages. For CLIC, detailed studies are available and have been presented in \cite{Kalinowski:2018kdn,deBlas:2018mhx,Zarnecki:2020nnw,Zarnecki:2020swm,Klamka:2728552}; we therefore omit their discussion here. {The main result for 3\,TeV CLIC is that the discovery reach for charged scalar pair-production extends to up to scalar masses of 1\,TeV.}
At 27 \TeV, we find that the range up to 1 \TeV~ can basically be covered in all channels, although some of the {BPs and} HPs still remain elusive in the $AA$ channel. At a 100 \TeV~ collider, the number of HPs that remain inaccessible in this channel decreases. Including again $AA$ production with additional jets, only two of the HP points remain inaccessible in this channel according to our simple counting criterium. {We also briefly comment on the possibility of using proton spectrometers at hadron colliders to tag processes induced via photon-photon fusion. At tree-level, only charged scalar pair-production is possible. Cross sections for these processes are much smaller than for direct pair-production, but some points are within range at both HL-LHC as well as a 100 \TeV collider assuming high tagging efficiency of forward proton spectrometers.}  

At a muon collider, we can again discuss both direct as well as VBF-like production channels. For direct production, $AH$ as well as $H^+H^-$ seem to be accessible at all center-of-mass energies considered for all BPs and HPs. For the VBF-like probes, with 10 \TeV~ center-of-mass energy, basically all low-mass BPs as well a subset of high-mass HPs are accessible for $AA$ production, which might provide an interesting cross check. This corresponds to a mass scale for $M_A$ in this channel of about 700 \GeV. At 30 \TeV, all channels should be accessible assuming target luminosity over the whole runtime. In addition, for almost all scenarios the VBF-induced production of $H^+H^-$ gives higher cross sections than direct pair-production, with the exception of the HPs at 10 \TeV. 

We again want to emphasize that our rough criterium needs to be supported by detailed studies for each scenario, including both signal and background. However, we consider the BPs and HPs presented here give useful guidelines for either phenomenological studies or experimental searches.

\section*{Acknowledgements}

This research was supported in parts by the National Science Centre,
Poland, the HARMONIA project under contract UMO-2015/18/M/ST2/00518
(2016-2021), OPUS project under contract UMO-2017/25/B/ST2/00496
(2018-2021), as well as the COST action CA16201 - Particleface. The work of TR was partially supported by grant K 125105 of
the National Research, Development and Innovation Fund in Hungary. JK thanks Gudrid Moortgat-Pick for her hospitality and the DFG for support 
through the SFB~676 ``Particles, Strings and the Early Universe'' 
during the initial stage of this project. {We also want to thank the \texttt{MicrOMEGAs} authors for useful discussions regarding different versions of their code.}
\begin{appendix}
\section{Benchmark tables using \texttt{micrOMEGAs$\_$5.2.4}}\label{app:newmo}

\begin{table}[tbp]
\begin{center}
  \small
  \begin{tabular}{|l|l|l|l|l|l||l|l|}
\hline
\multirow{2}{*}{No.} & \multirow{2}{*}{$M_H$} & \multirow{2}{*}{$M_A$} & \multirow{2}{*}{$M_{H^\pm}$} & \multirow{2}{*}{$\lambda_{345}$}& \multirow{2}{*}{$\Omega_H h^2$} & \multirow{2}{*}{$\Omega_H h^2$} & \multirow{2}{*}{$\Omega_H h^2$}\\
[+1mm] 
& & & & & \texttt{5.0.4}   & \texttt{5.2.4} & \texttt{5.0.4}, \texttt{fast=0} \\ \hline
\sl{BP1} & 72.77 & 107.803 & 114.639 & -0.00440723&0.11998 & 0.12277&0.12245\\\hline
BP2 & 65 & 71.525 & 112.85 & 0.0004& 0.07076 & 0.07085 &0.070725\\ \hline
BP3 & 67.07 & 73.222 & 96.73 & 0.00738&0.06159 & 0.061706&0.061569\\ \hline
BP4 & 73.68 & 100.112 & 145.728 &  -0.00440723 & 0.089114& 0.08854& 0.088507\\ \hline
\textbf{BP6} & 72.14 & 109.548 & 154.761 & -0.00234 & 0.117& 0.12155&0.12104\\ \hline
BP7 & 76.55 & 134.563 & 174.367 &  0.0044 & 0.031381& 0.031334&0.031303\\ \hline
\sl{BP8} & 70.91 & 148.664 & 175.89 &  0.0058&0.12207 & 0.122618& 0.12537\\ \hline
BP9 & 56.78 & 166.22 & 178.24 & 0.00338 & 0.081243& 0.081226& 0.080547\\ \hline
BP23 & 62.69 & 162.397 & 190.822 &  0.0056& 0.065 & 0.0608& 0.062146\\ \hline
BP10 & 76.69 & 154.579 & 163.045 &  0.0096 & 0.028125& 0.028312& 0.028293\\ \hline
BP11 & 98.88 & 155.037 & 155.438 &  -0.0628 &0.002735& 0.002735&0.0027319\\ \hline
BP12 & 58.31 & 171.148 & 172.96 & 0.00762 &0.0064104& 0.0064098&0.0063558\\ \hline
BP13 & 99.65 & 138.484 & 181.321 &  0.0532 &0.0012541& 0.0012541&0.0012534\\ \hline
\sl{BP14} & 71.03 & 165.604 & 175.971 & 0.00596 &0.11833& 0.12245&0.12169\\ \hline
\sl{BP15} & 71.03 & 217.656 & 218.738 & 0.00214& 0.12217 & 0.12647&0.12571\\ \hline
\sl{BP16} & 71.33 & 203.796 & 229.092 & -0.00122& 0.12205 & 0.12686&0.12615\\ \hline
BP18 & 147 & 194.647 & 197.403 & -0.018 & 0.0017711& 0.0017711& 0.0017667\\ \hline
BP19 & 165.8 & 190.082 & 195.999 &  -0.004 &0.0028308 & 0.0028308&0.0028209\\ \hline
BP20 & 191.8 & 198.376 & 199.721 & 0.008 & 0.0084219& 0.0084219&0.0083755\\ \hline
\textbf{BP21} & 57.475 & 288.031 & 299.536 & 0.00192& 0.11942& 0.11937& 0.11845\\ \hline
\sl{BP22} & 71.42 & 247.224 & 258.382 &  -0.0032 & 0.12206& 0.12699&0.1263\\ \hline
\end{tabular}

\caption{ As table \ref{tab:bench} (without $\lam_2$ and on- or off-shell information for intermediate gauge bosons), with dark matter relic density calculated using \texttt{micrOMEGAs$\_$5.2.4} and  \texttt{micrOMEGAs$\_$5.0.4} with \texttt{fast=0}. {Note that several benchmark points, selected previously to match PLANCK measurements, result in relic density slightly above the assumed limit (indicated by slashed font).}
\label{tab:bench_nmo}
}
\end{center}
\end{table}

We present the benchmark points from table \ref{tab:bench}, were  \texttt{micrOMEGAs$\_$5.2.4} has been used in the relic density calculation, in table \ref{tab:bench_nmo}. For selected benchmark points, deviations can be up to $7\%$.  We also present values for \texttt{micrOMEGAs$\_$5.0.4} for \texttt{fast=0} in the integration setup (see \cite{Belanger:2018mqt} for details). 
\section{Diagrams contributing to $\mu^+\,\mu^-\,\rightarrow\,H^+\,H^-\,\nu_\mu\,\bar{\nu}_\mu$}\label{app:diags}
\subsection{Diagrams via $W^+ W^-$ fusion (GI I)}
\begin{center}
\begin{figure}[h!]
\centering
\includegraphics[width=0.9\textwidth]{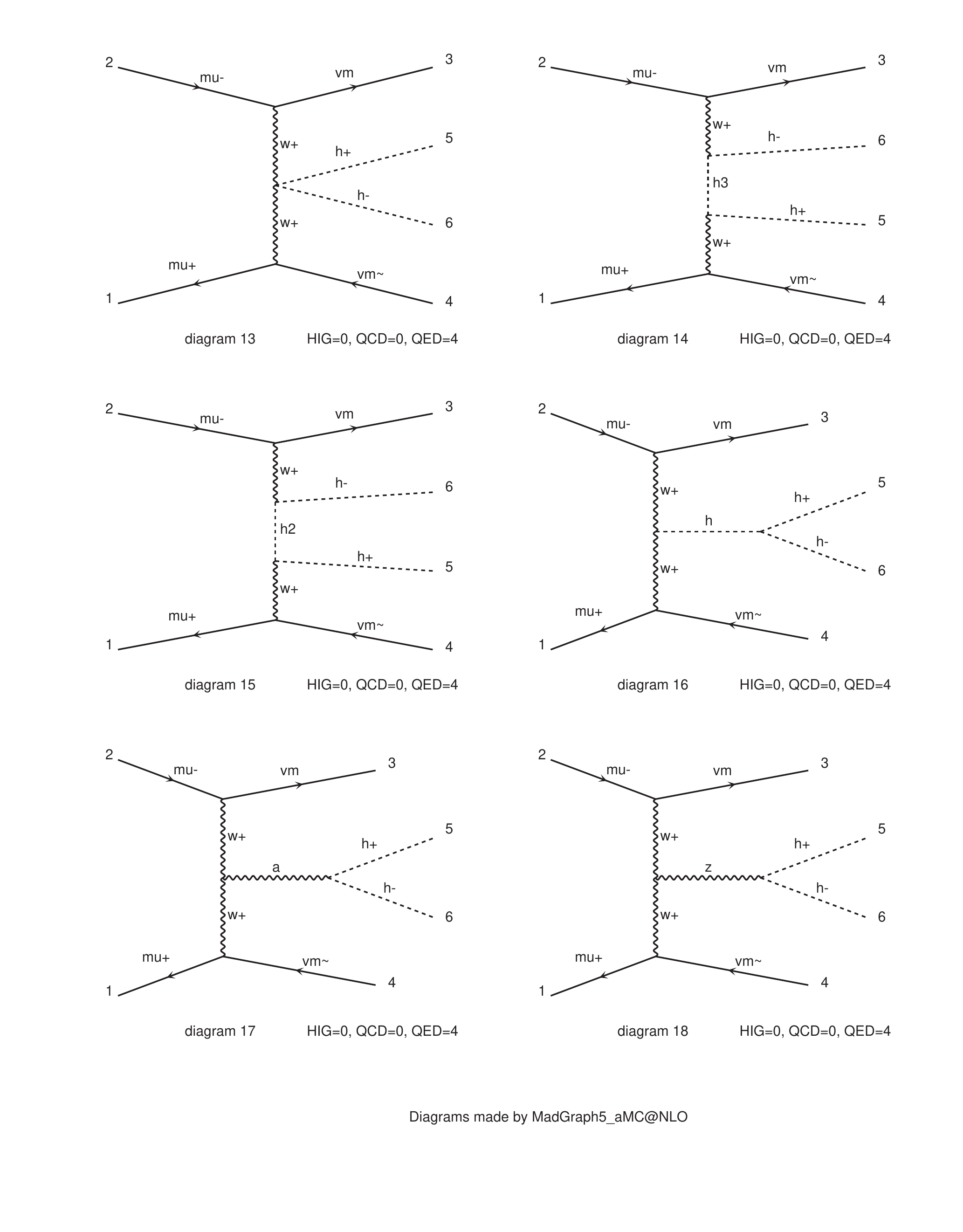}
\caption{GI I diagrams for $W^+ W^-$ fusion.}
\end{figure}
\end{center}
\newpage
\subsection{Sample diagrams via $W \mu $ fusion (GI II)}
\begin{center}
\begin{figure}[h!]
\begin{center}
\includegraphics[width=0.90\textwidth]{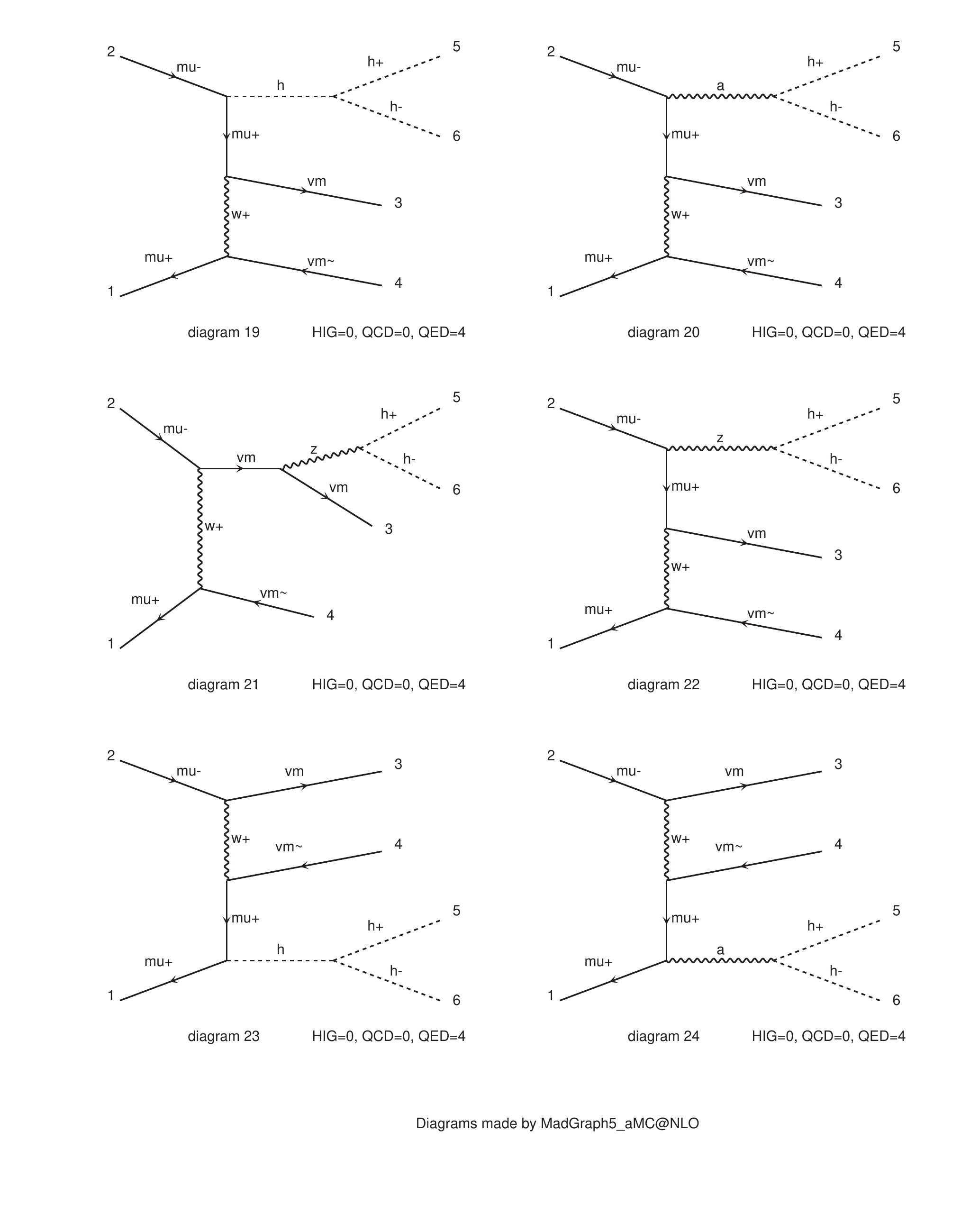}
\end{center}
\caption{GI II diagrams for $W \mu$ fusion.} 
\end{figure}
\end{center}
\newpage
\section{Sample diagrams contributing to $\mu^+\,\mu^-\,\rightarrow\,A\,A\,\nu_\mu\,\bar{\nu}_\mu$}\label{app:diagsaa}
\begin{center}
\begin{figure}[h!]
\begin{center}
\includegraphics[width=0.90\textwidth]{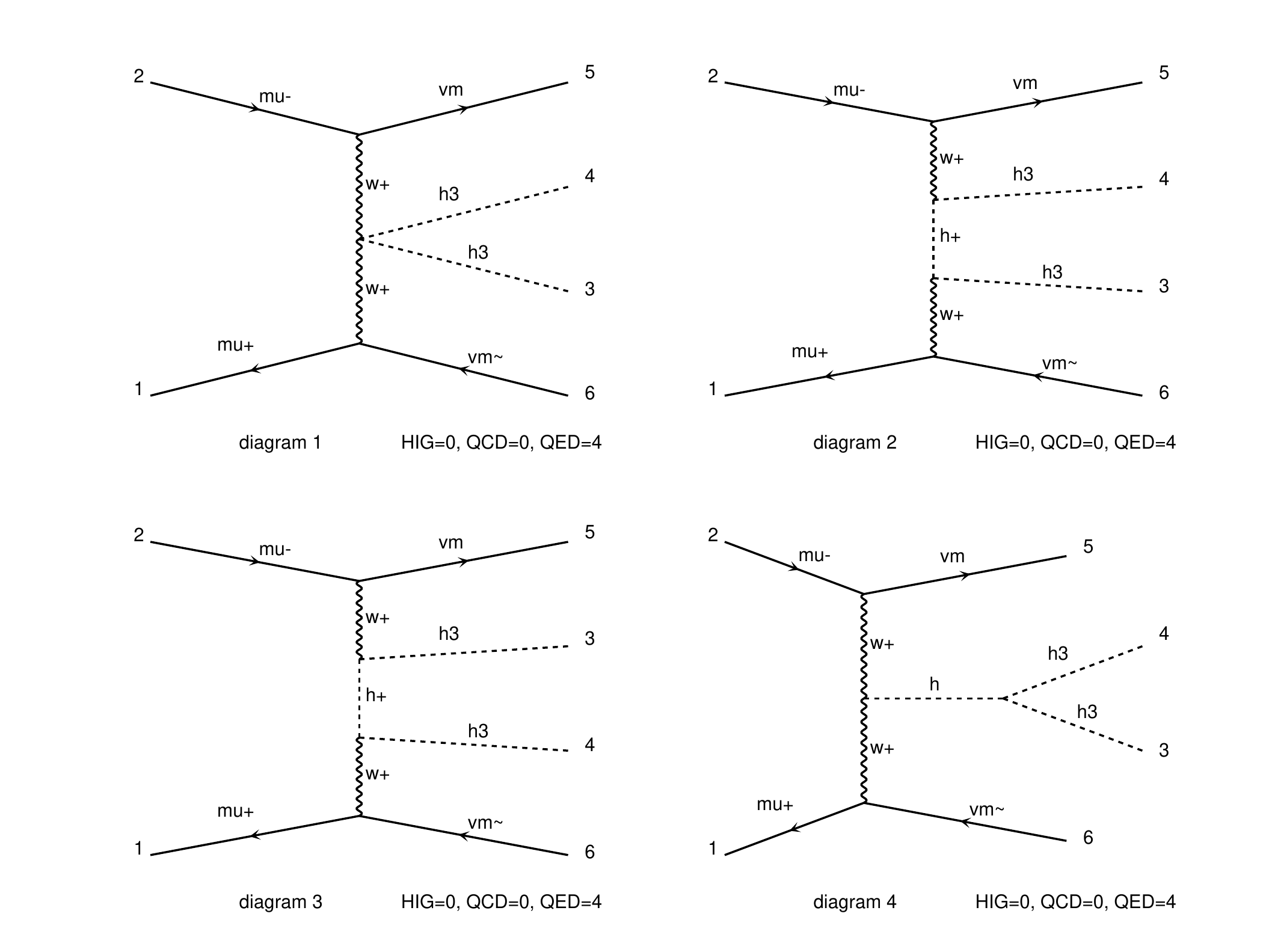}
\end{center}
\caption{Diagrams for $W^+ W^-$ fusion.}
\end{figure}
\end{center}
\end{appendix}
\newpage
\bibliographystyle{hunsrt}


\end{document}